\documentclass[journal]{IEEEtran}

\usepackage{cite}
\usepackage{amsmath,amssymb,amsfonts}
\usepackage{multirow}
\usepackage{graphicx,import}
\usepackage{xcolor}
\usepackage{siunitx}
\usepackage{float}
\usepackage[caption=false,font=footnotesize]{subfig}
\usepackage{tikz}
\usepackage{pgfplots}
\usepackage{multicol}
\usepackage{booktabs}
\usepackage[utf8]{inputenc}
\usepgfplotslibrary{fillbetween}

\usepackage{tabularx}
\pgfplotsset{compat=1.18}

\renewcommand{\Re}[1]{\mathrm{Re}\left\{#1\right\}}

\newcommand{\E}[2][]
{
	\ifthenelse{\isempty{#1}}
		{\mathbb{E}\left[#2\right]}
		{\mathbb{E}_{#1}\left[#2\right]}
}

\newcommand{\var}[2][]
{
	\ifthenelse{\isempty{#1}}
		{\text{Var}\left[#2\right]}
		{\text{Var}_{#1}\left[#2\right]}
}
\newcommand{\cov}[2][]
{
	\ifthenelse{\isempty{#1}}
		{\text{Cov}\left[#2\right]}
		{\text{Cov}_{#1}\left[#2\right]}
}
\newcommand{\rpm}{\sbox0{$1$}\sbox2{$\scriptstyle\pm$}
  \raise\dimexpr(\ht0-\ht2)/2\relax\box2 }

\usepackage[acronym,nomain,nonumberlist,nopostdot,xindy]{glossaries}

\newglossary[nlg]{notation}{not}{ntn}{Notation}

\setacronymstyle{long-short}

\setglossarystyle{super}

\glsnoexpandfields

\newacronym{3D}{3D}{three-dimensional}
\newacronym{3GPP}{3GPP}{3rd-Generation Partnership Project}
\newacronym{1G}{1G}{first generation}
\newacronym{2G}{2G}{second generation}
\newacronym{3G}{3G}{third generation}
\newacronym{4G}{4G}{fourth generation}
\newacronym{5G}{5G}{fifth generation}
\newacronym{6G}{6G}{sixth generation}
\newacronym{802.11}{802.11}{IEEE 802.11 specifications}

\newacronym{ACM}{ACM}{adaptive coded modulation}
\newacronym{ADC}{ADC}{analog-to-digital converter}  
\newacronym{ASIC}{ASIC}{Application-Specific Integrated Circuit}
\newacronym{ASP}{ASP}{Application Service Provider}
\newacronym[longplural={Action Sequences}]{ASQ}{ASQ}{Action Sequence}
\newacronym{AVB}{AVB}{Audio-Video Bridging}
\newacronym{AXIS}{AXIS}{AXI Stream}
\newacronym{A/D}{A/D}{analog-to-digital}
\newacronym{ABF}{ABF}{analog beamforming}
\newacronym{AM}{AM}{amplitude modulation}
\newacronym{AP}{AM}{access point}
\newacronym{AR}{AR}{augmented reality}
\newacronym{ASK}{ASK}{amplitude-shift keying}
\newacronym{ASIP}{ASIP}{Application Specific Integrated Processors}
\newacronym{AWGN}{AWGN}{additive white Gaussian noise}
\newacronym{AI}{AI}{artificial intelligence}
\newacronym{AGC}{AGC}{automatic gain control}
\newacronym{AGV}{AGV}{automated guided vehicles}
\newacronym{AN}{AN}{actuator node}

\newacronym{BCJR}{BCJR}{Bahl, Cocke, Jelinek and Raviv}
\newacronym{BER}{BER}{bit error rate}
\newacronym{BFDM}{BFDM}{bi-orthogonal frequency division multiplexing}
\newacronym{BPSK}{BPSK}{binary phase shift keying}
\newacronym{BS}{BS}{base station} 
\newacronym{BW}{BW}{bandwidth}
\newacronym{CA}{CA}{carrier aggregation}
\newacronym{CAF}{CAF}{cyclic autocorrelation function}
\newacronym{Car-2-x}{Car-2-x}{car-to-car and car-to-infrastructure communication}
\newacronym{CAZAC}{CAZAC}{constant amplitude zero autocorrelation waveform}
\newacronym{CB-FMT}{CB-FMT}{cyclic block filtered multitone}
\newacronym{CCDF}{CCDF}{complementary cumulative density function}
\newacronym{CDF}{CDF}{cumulative density function}
\newacronym{CDMA}{CDMA}{code-division multiple access}
\newacronym{CMOS}{CMOS}{complementary metal-oxide-semiconductor}
\newacronym{CFO}{CFO}{carrier frequency offset}
\newacronym{CIR}{CIR}{channel impulse response}
\newacronym{CM}{CM}{complex multiplication}
\newacronym{COFDM}{COFDM}{coded-\gls{OFDM}}
\newacronym{CoMP}{CoMP}{coordinated multi point}
\newacronym{COQAM}{COQAM}{cyclic OQAM}
\newacronym{COTS}{COTS}{commercial off-the-shelf}
\newacronym{CP}{CP}{cyclic prefix}
\newacronym{CR}{CR}{cognitive radio}
\newacronym{CRC}{CRC}{cyclic redundancy check}
\newacronym{CRLB}{CRLB}{Cram\'{e}r-Rao lower bound}
\newacronym{CS}{CS}{cyclic suffix}
\newacronym{CSI}{CSI}{channel state information}
\newacronym{CSMA}{CSMA}{carrier-sense multiple access}
\newacronym{CWCU}{CWCU}{component-wise conditionally unbiased}
\newacronym{CU}{CU}{central unit}
\newacronym{CN}{CN}{control node}
\newacronym{CPS}{CPS}{cyber-physical system}
\newacronym{D/A}{D/A}{digital-to-analog}
\newacronym{D2D}{D2D}{device-to-device}
\newacronym{DAC}{DAC}{digital-to-analog converter}
\newacronym{DBF}{DBF}{digital beamforming}
\newacronym{DC}{DC}{direct current}
\newacronym{DFE}{DFE}{decision feedback equalizer}
\newacronym{DFT}{DFT}{discrete Fourier transform}
\newacronym{DL}{DL}{downlink}
\newacronym{DMT}{DMT}{discrete multitone}
\newacronym{DNN}{DNN}{deep neural network}
\newacronym{DoA}{DoA}{direction of arrival}
\newacronym{DSA}{DSA}{dynamic spectrum access}
\newacronym{DSL}{DSL}{digital subscriber line}
\newacronym{DSP}{DSP}{digital signal processor}
\newacronym{DTFT}{DTFT}{discrete-time Fourier transform}
\newacronym{DVB}{DVB}{digital video broadcasting}
\newacronym{DVB-T}{DVB-T}{terrestrial digital video broadcasting}
\newacronym{DWMT}{DWMT}{discrete wavelet multi tone}
\newacronym{DZT}{DZT}{discrete Zak transform}
\newacronym{E2E}{E2E}{end-to-end}
\newacronym{eNodeB}{eNodeB}{evolved node b base station}
\newacronym{E-SNR}{E-SNR}{effective signal-to-noise ratio}
\newacronym{EVD}{EVD}{eigenvalue decomposition}
\newacronym{ERP}{ERP}{effective radiated power}
\newacronym{FBMC}{FBMC}{filter bank multicarrier}
\newacronym{FD}{FD}{frequency-domain}
\newacronym{FDD}{FDD}{frequency-division duplexing}
\newacronym{FDE}{FDE}{frequency domain equalization}
\newacronym{FDM}{FDM}{frequency division multiplex}
\newacronym{FDMA}{FDMA}{frequency-division multiple access}
\newacronym{FEC}{FEC}{forward error correction}
\newacronym{FER}{FER}{frame error rate}
\newacronym{FFT}{FFT}{fast Fourier transform}
\newacronym{FIR}{FIR}{finite impulse response}
\newacronym{FM}{FM}{frequency modulation}
\newacronym{FMT}{FMT}{filtered multi tone}
\newacronym{FO}{FO}{frequency offset}
\newacronym{F-OFDM}{F-OFDM}{filtered-\acs{OFDM}}
\newacronym{FPGA}{FPGA}{field programmable gate array}
\newacronym{FSC}{FSC}{frequency selective channel}
\newacronym{FS-OQAM-GFDM}{FS-OQAM-GFDM}{frequency-shift OQAM-GFDM}
\newacronym{FSPL}{FSPL}{free-space path loss}
\newacronym{FT}{FT}{Fourier transform}
\newacronym{FTD}{FTD}{fractional time delay}
\newacronym{FTN}{FTN}{faster-than-Nyquist}
\newacronym{FOM}{FOM}{figure-of-merit}

\newacronym{GFDM}{GFDM}{generalized frequency division multiplexing}
\newacronym{GFDMA}{GFDMA}{generalized frequency division multiple access}
\newacronym{GMC-CDM}{GMC-CDM}{generalized	multicarrier code-division multiplexing}
\newacronym{GNSS}{GNSS}{global navigation satellite system}
\newacronym{GS}{GS}{guard symbols}
\newacronym{GSM}{GSM}{Groupe Sp\'{e}cial Mobile}
\newacronym{GUI}{GUI}{graphical user interface}
\newacronym{GN}{GN}{gateway node}

\newacronym{H2H}{H2H}{human-to-human}
\newacronym{H2M}{H2M}{human-to-machine}
\newacronym{HARQ}{HARQ}{hybrid automatic repeat request}
\newacronym{HBF}{HBF}{hybrid beamforming}
\newacronym{HN}{HN}{haptic node}
\newacronym{HPBW}{HPBW}{half power beam width}
\newacronym{HSI}{HSI}{human system interface}
\newacronym{HMI}{HMI}{human machine interface}
\newacronym{HTC}{HTC}{human type communication}

\newacronym{I}{I}{in-phase}
\newacronym{i.i.d.}{i.i.d.}{independent and identically distributed}
\newacronym{IB}{IB}{in-band}
\newacronym{IBI}{IBI}{inter-block interference}
\newacronym{IC}{IC}{interference cancellation}
\newacronym{ICI}{ICI}{inter-carrier interference}
\newacronym{ICT}{ICT}{information and communication technologies}
\newacronym{ICV}{ICV}{information coefficient vector}
\newacronym{IDFT}{IDFT}{inverse discrete Fourier transform}
\newacronym{IDMA}{IDMA}{interleave division multiple access}
\newacronym{IEEE}{IEEE}{institute of electrical and electronics engineers}
\newacronym{IF}{IF}{intermediate frequency}
\newacronym{IFFT}{IFFT}{inverse fast Fourier transform}
\newacronym{IoT}{IoT}{Internet of Things}
\newacronym{IOTA}{IOTA}{isotropic orthogonal transform algorithm}
\newacronym{IP}{IP}{internet protocole}
\newacronym{IP-core}{IP-core}{intellectual property core}
\newacronym{IR-UWB}{IR-UWB}{impulse-radio ultra-wideband}
\newacronym{ISDB-T}{ISDB-T}{terrestrial integrated services digital broadcasting}
\newacronym{ISDN}{ISDN}{integrated services digital network}
\newacronym{ISI}{ISI}{inter-symbol interference}
\newacronym{ITU}{ITU}{International Telecommunication Union}
\newacronym{IUI}{IUI}{inter-user interference}
\newacronym{IMT}{IMT}{ international mobile telecommunications}
\newacronym{IR}{IR}{impulse radio}

\newacronym{K}{K}{Key Technologies and Methods}
\newacronym{KPI}{KPI}{Key Performance Indicator}

\newacronym{LAN}{LAN}{local area netwrok}
\newacronym{LDPC}{LDPC}{low-density parity check}
\newacronym{LLMS}{LLMS}{linear least-mean square}
\newacronym{LLR}{LLR}{log-likelihood ratio}
\newacronym{LMMSE}{LMMSE}{linear minimum mean square error}
\newacronym{LNA}{LNA}{low noise amplifier}
\newacronym{LO}{LO}{local oscillator}
\newacronym{LOS}{LOS}{line-of-sight}
\newacronym{LoS}{LoS}{line of sight}
\newacronym{LP}{LP}{low-pass}
\newacronym{LPF}{LPF}{low-pass filter}
\newacronym{LS}{LS}{least squares}
\newacronym{LTE}{LTE}{long term evolution}
\newacronym{LTE-A}{LTE-A}{LTE-Advanced}
\newacronym{LTIV}{LTIV}{linear time invariant}
\newacronym{LTV}{LTV}{linear time variant}
\newacronym{LUT}{LUT}{lookup table}

\newacronym{M2M}{M2M}{machine-to-machine}
\newacronym{MA}{MA}{multiple access}
\newacronym{MAC}{MAC}{multiple access control}
\newacronym{MAP}{MAP}{maximum a posteriori}
\newacronym{MC}{MC}{multicarrier}
\newacronym{MCA}{MCA}{multicarrier access}
\newacronym{MCM}{MCM}{multicarrier modulation}
\newacronym{MCS}{MCS}{modulation coding scheme}
\newacronym{MF}{MF}{matched filter}
\newacronym{MF-SIC}{MF-SIC}{matched filter with successive interference cancellation}
\newacronym{MIMO}{MIMO}{multiple-input multiple-output}
\newacronym{MISO}{MISO}{multiple-input single-output}
\newacronym{ML}{ML}{machine learning}
\newacronym{MLD}{MLD}{maximum likelihood detection}
\newacronym{MLE}{MLE}{maximum likelihood estimator}
\newacronym{MMSE}{MMSE}{minimum mean squared error}
\newacronym{mmwave}{mmWave}{millimeter-wave}
\newacronym{MRC}{MRC}{maximum ratio combining}
\newacronym{MS}{MS}{mobile stations}
\newacronym{MSE}{MSE}{mean squared error}
\newacronym{MSK}{MSK}{Minimum-shift keying}
\newacronym{MOS}{MOS}{metal-oxide-semiconductor}
\newacronym{MSSS}{MSSS}{mean-square signal separation}
\newacronym{MTC}{MTC}{machine type communication}
\newacronym{MU}{MU}{multi-user}
\newacronym{MVUE}{MVUE}{minimum variance unbiased estimator}
\newacronym{MVDR}{MVDR}{minimum variance distortionless response}
\newacronym{MEC}{MEC}{multiaccess edge cloud}

\newacronym{NEF}{NEF}{noise enhancement factor}
\newacronym{NLOS}{NLOS}{non-line-of-sight}
\newacronym{NMSE}{NMSE}{normalized mean-squared error}
\newacronym{NOMA}{NOMA}{non-orthogonal multiple access}
\newacronym{NPR}{NPR}{near-perfect reconstruction}
\newacronym{NRZ}{NRZ}{non-return-to-zero}
\newacronym{NC}{NC}{network controller}
\newacronym{NFV}{NFV}{network function virtualization}
\newacronym{NR}{NR}{new radio}

\newacronym{OFDM}{OFDM}{orthogonal frequency division multiplexing}
\newacronym{OFDMA}{OFDMA}{orthogonal frequency division multiple access}
\newacronym{OOB}{OOB}{out-of-band}
\newacronym{OOK}{OOK}{on-off keying}
\newacronym{OQAM}{OQAM}{offset quadrature amplitude modulation}
\newacronym{OQPSK}{OQPSK}{offset quadrature phase shift keying}
\newacronym{OTFS}{OTFS}{orthogonal time frequency space}

\newacronym{PA}{PA}{power amplifier}
\newacronym{PAA}{PAA}{phased antenna array}
\newacronym{PAM}{PAM}{pulse amplitude modulation}
\newacronym{PAPR}{PAPR}{peak-to-average power ratio}
\newacronym{PC-CC}{PC-CC}{parallel concatenated convolutional code}
\newacronym{PC-PAS}{PC-PAS}{polar-coded probabilistic amplitude shaping}
\newacronym{PCB}{PCB}{printed circuit board}
\newacronym{PCP}{PCP}{pseudo-circular pre/post-amble}
\newacronym{PD}{PD}{probability of detection}
\newacronym{pdf}{pdf}{probability density function}
\newacronym{PDF}{PDF}{probability distribution function}
\newacronym{PDP}{PDP}{power delay profile}
\newacronym{PFA}{PFA}{probability of false alarm}
\newacronym{PHY}{PHY}{physical layer}
\newacronym{PIC}{PIC}{parallel interference cancellation}
\newacronym{PLC}{PLC}{power line communication}
\newacronym{PMF}{PMF}{probability mass function}
\newacronym{PN}{PN}{phase noise}
\newacronym{ppm}{ppm}{parts per million}
\newacronym{PPM}{PPM}{pulse position modulation}
\newacronym{PRB}{PRB}{physical resource block}
\newacronym{PRBS}{PRBS}{pseudorandom binary sequence}
\newacronym{PSD}{PSD}{power spectral density}
\newacronym{PAE}{PAE}{power added efficiency}
\newacronym{PMEPR}{PMEPR}{peak-to-mean-envelope power ratio}
\newacronym{PLL}{PLL}{phase-locked loop}
\newacronym{PFD}{PFD}{phase/frequency detector}

\newacronym{Q}{Q}{quadrature-phase}
\newacronym{QAM}{QAM}{quadrature amplitude modulation}
\newacronym{QoS}{QoS}{quality of service}
\newacronym{QPSK}{QPSK}{quadrature phase shift keying}
\newacronym{R/W}{R/W}{read-or-write}

\newacronym{RAM}{RAM}{random-access memmory}
\newacronym{RAN}{RAN}{radio access network}
\newacronym{RAT}{RAT}{radio access technologies}
\newacronym{RC}{RC}{raised-cosine}
\newacronym{RCUB}{RCUB}{random coding union bound}
\newacronym{RF}{RF}{radio frequency}
\newacronym{rms}{rms}{root mean square}
\newacronym{RRC}{RRC}{root-raised-cosine}
\newacronym{RW}{RW}{read-and-write}
\newacronym{Rx}{RX}{receiver}
\newacronym{RLL}{RLL}{runlength limited}
\newacronym{RU}{RU}{radio unit}
\newacronym{RSSI}{RSSI}{receive signal strength indicator}

\newacronym{SAW}{SAW}{surface acoustic wave}
\newacronym{SC}{SC}{successive cancellation}
\newacronym{SCA}{SCA}{single-carrier access}
\newacronym{SC-FDE}{SC-FDE}{single-carrier with frequency domain equalization}
\newacronym{SC-FDM}{SC-FDM}{single-carrier frequency division multiplexing}
\newacronym{SC-FDMA}{SC-FDMA}{single-carrier frequency division multiple access}
\newacronym{SD}{SD}{sphere decoding}
\newacronym{SDD}{SDD}{space-division duplexing}
\newacronym{SDMA}{SDMA}{space division multiple access}
\newacronym{SDR}{SDR}{software-defined radio}
\newacronym{SDW}{SDW}{software-defined waveform}
\newacronym{SEFDM}{SEFDM}{spectrally efficient frequency division multiplexing}
\newacronym{SE-FDM}{SE-FDM}{spectrally efficient frequency division multiplexing}
\newacronym{SER}{SER}{symbol error rate}
\newacronym{SIC}{SIC}{successive interference cancellation}
\newacronym{SINR}{SINR}{signal-to-interference-plus-noise ratio}
\newacronym{SIR}{SIR}{signal-to-interference ratio}
\newacronym{SISO}{SISO}{single-input single-output}
\newacronym{SITO}{SITO}{switched injection-triggered oscillator}
\newacronym{SMS}{SMS}{Short Message Service}
\newacronym{SNR}{SNR}{signal-to-noise ratio}
\newacronym{SRR}{SRR}{super-regenerative receiver}
\newacronym{STC}{STC}{space-time coding}
\newacronym{STFT}{STFT}{short-time Fourier transform}
\newacronym{STO}{STO}{symbol time offset}
\newacronym{SU}{SU}{single user}
\newacronym{SVD}{SVD}{singular value decomposition}
\newacronym{SCL}{SCL}{successive cancellation list}
\newacronym{SDN}{SDN}{software defined networking}
\newacronym{SE}{SE}{spectral efficiency}
\newacronym{SN}{SN}{sensor node}

\newacronym{TDM}{TDM}{time division multiplex}
\newacronym{TD-ZXM}{TD-ZXM}{time-derived zero-crossing modulation}
\newacronym{TDD}{TDD}{time-division duplexing}
\newacronym{TDMA}{TDMA}{time-division multiple access}
\newacronym{TFL}{TFL}{time-frequency localization}
\newacronym{THz}{THz}{terahertz}
\newacronym{TO}{TO}{time offset}
\newacronym{TS-OQAM-GFDM}{TS-OQAM-GFDM}{time-shifted OQAM-GFDM}
\newacronym{Tx}{Tx}{transmitter}
\newacronym{TI}{TI}{tactile Internet}
\newacronym{TSM}{TSM}{tactile service manager}
\newacronym{TSN}{TSN}{Time-Sensitive Networking}
\newacronym{TTI}{TTI}{transmission time interval}

\newacronym{UE}{UE}{user equipment}
\newacronym{UFMC}{UFMC}{universally filtered multicarrier}
\newacronym{UL}{UL}{uplink}
\newacronym{ula}{ULA}{uniform linear array}
\newacronym{US}{US}{uncorrelated scattering}
\newacronym{USB}{USB}{universal serial bus}
\newacronym{UW}{UW}{unique word}
\newacronym{URLLC}{URLLC}{ultra-reliable low-latency communications}

\newacronym{VCO}{VCO}{voltage-controlled oscillator}

\newacronym{WCP}{WCP}{windowing and \acs{CP}}	
\newacronym{WHT}{WHT}{Walsh-Hadamard transform}
\newacronym{WiMAX}{WiMAX}{worldwide interoperability for microwave access}
\newacronym{WLAN}{WLAN}{wireless local area network}
\newacronym{W-OFDM}{W-OFDM}{windowed-\acs{OFDM}}	
\newacronym{WOLA}{WOLA}{windowing and overlapping}	
\newacronym{WuRx}{WuRx}{wakeup receiver}
\newacronym{WSS}{WSS}{wide-sense stationary}

\newacronym{ZoH}{ZoH}{Zero-Order-Hold}
\newacronym{ZXM}{ZXM}{zero-crossing modulation}
\newacronym{ZCT}{ZCT}{Zadoff-Chu transform}
\newacronym{ZF}{ZF}{zero-forcing}
\newacronym{ZMCSCG}{ZMCSCG}{zero-mean circularly-symmetric complex Gaussian}
\newacronym{ZP}{ZP}{zero-padding}
\newacronym{ZT}{ZT}{zero-tail}

\newglossaryentry{pi}{%
  type=notation,
  name={\ensuremath{\pi}},
  description={ratio of circumference of a circle to its diameter},
  sort={pi}
}

\definecolor{mycolor2}{rgb}{0.00000,0.19600,0.42500}
\definecolor{mycolor1}{rgb}{0.90900,0.41100,0.2000}
\definecolor{mycolor3}{rgb}{0.4660 0.6740 0.1880}
\definecolor{mycolor4}{rgb}{0.55400,0.3300,0.62600}
\definecolor{mycolor5}{rgb}{0.55600,0.55600,0.55600}
\definecolor{mycolorgrey}{rgb}{0.55600,0.55600,0.55600}
\definecolor{mycolorblack}{rgb}{0,0,0}

\definecolor{mycolor5}{rgb}{1,0.8431372549019608,0}%
\definecolor{mycolor6}{rgb}{0.117647058, 0.5647058823529412, 1}%
\definecolor{mycolor7}{rgb}{0.0, 0.39215686, 0}%
\definecolor{mycolor8}{rgb}{1,0.07843137,0.5764705882352941}%
\definecolor{mycolor9}{rgb}{0.55900,0.21100,0.0000}

\pgfplotsset{every axis plot/.append style={line width=1pt}}

\newcommand{\smallFigHeight}{0.9in}
\begin{document}
\title{Energy-Efficient Mobile Communications using an Adaptive Gearbox-PHY under Hardware Constraints}
\author{Florian Gast, \textit{Member, IEEE}, Meik Dörpinghaus, \textit{Member, IEEE}, and Gerhard P. Fettweis, \textit{Fellow, IEEE}
\thanks{This work was presented in part at the International Symposium on Wireless Communication Systems (ISWCS) in Rio de Janeiro, Brazil, in 2024\cite{Gast202407} and at  the  Wireless Communications and Networking Conference (WCNC) in Milan, Italy, in 2025 \cite{Gast202503}.}
\thanks{F. Gast is with the Barkhausen Institut, Dresden, Germany, E-mail: florian.gast@barkhauseninstitut.org.}
\thanks{M. Dörpinghaus is with the Vodafone Chair Mobile Communications Systems, Technische Universität Dresden, Germany, E-mail: {meik.doerpinghaus}@tu-dresden.de}
\thanks{G. Fettweis is with both the Barkhausen Institut and the Vodafone Chair Mobile Communications Systems, Technische Universität Dresden, Germany, E-mail: {gerhard.fettweis@barkhauseninstitut.org}}
\thanks{This work was supported by the German Federal Ministry of Education and
Research (BMBF) (6G-life, project-ID 16KISK001K) and the Saxon State government out of the state budget approved by the Saxon State Parliament.}
}

\makeatletter
\def\ps@IEEEtitlepagestyle{%
  \def\@oddfoot{\mycopyrightnotice}%
  \def\@evenfoot{}%
}
\def\mycopyrightnotice{%
  {\small
  \parbox{\textwidth}{\centering
  This work has been submitted to the IEEE for possible publication.
  Copyright may be transferred without notice, after which this version may no longer be accessible.}}%
}
\makeatother

\maketitle

\begin{abstract}
Future mobile networks must achieve substantial improvements in energy efficiency to offset the anticipated traffic growth. Despite this requirement, many discussions regarding physical layer design remain primarily focused on peak data rates and spectral efficiency, even though typical network operation is dominated by low-data-rate regimes.
To address this mismatch, the Gearbox-PHY was proposed as an energy-efficient physical layer architecture that dynamically switches between modulation schemes and their associated analog front ends in order to adapt to varying operating requirements.
This paper quantifies the achievable energy savings by jointly modeling front end power consumption and hardware-aware spectral efficiency to formulate an energy-per-bit minimization problem. To move beyond idealized assumptions, non-ideal hardware effects, including oscillator phase noise and limited quantizer resolution, are incorporated. These impairments simultaneously affect power consumption and achievable spectral efficiency, thereby introducing trade-offs between front end complexity, hardware non-linearities, spectral efficiency, and energy efficiency. 
Numerical results demonstrate that the Gearbox-PHY enables significant energy savings, particularly at low data rates. Evaluations with spatially distributed users confirm that gains of up to two orders of magnitude persist in a cellular deployment scenario.
\end{abstract}
\begin{IEEEkeywords}
wireless communications, energy efficiency
\end{IEEEkeywords}
\section{Introduction}
The vision for \gls{6G} wireless networks extends beyond simply improving speed and capacity, aiming instead for significant advancements in industry and society. In this context, three key strategic drivers have been identified to shape the evolution of future mobile networks \cite{NGMN2021drivers}:

\begin{enumerate}
\item {Societal goals}, particularly addressing the United Nations sustainable development goals,
\item {Operational necessities}, emphasizing more efficient network operation and management, and
\item {Market expectations}, focusing on enabling new services and capabilities in a cost-effective manner.
\end{enumerate}
Energy efficiency is central to these drivers and directly impacts environmental sustainability, economic viability, and the feasibility of new applications.

Regarding the societal goals, the environmental impact of communications poses a necessity for improving energy efficiency in \gls{6G}. Due to their increasing energy demands, mobile networks substantially contribute to global greenhouse gas emissions. In 2015, global mobile network operations consumed approximately \SI{137}{TWh}, about $0.6\%$ of global electricity usage \cite{malmodin2018energy, GSMA2019}. This consumption rose significantly, by nearly $61\%$ in just five years, equivalent to an annual growth rate of around $10\%$ \cite{malmodin2018energy}. Similar trends have been reported in Germany, where mobile network energy consumption is expected to rise by $12\%$ annually until 2030 \cite{bles_umweltbezogene_nodate}. This rise in energy consumption also leads to significant $\text{CO}_2$ emissions of cellular communications, which reached about \SI{220}{Mt CO_2 e} globally by 2021 \cite{GSMA2021}. Without substantial improvements in energy efficiency, meeting the climate action goals of the United Nations becomes considerably more challenging.

From an economic perspective, reducing energy consumption is equally critical. An analysis from 2021 shows that energy costs account for $20\%$ to $40\%$ of the operational expenditures for \textit{typical} mobile network operators \cite{GSMA2021_SustainableTelco}. Considering that operational expenses represent roughly $37\%$ of total cost of ownership \cite[Section 4.6.2]{5GGuide}, energy-related expenses alone constitute nearly $15\%$ of the total ownership cost. Therefore, improving energy efficiency is essential for economically sustainable and competitive network operations.

Lastly, improved energy efficiency also facilitates new applications, particularly those involving battery-powered or energy-harvesting devices \cite{sharma2013issues}. For instance, wireless sensor networks deployed in remote or challenging environments require ultra-low-power operation to be viable.

\begin{figure}
\centering
\resizebox{0.35\textwidth}{!}{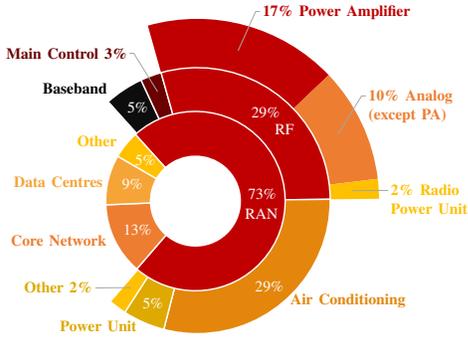}
\caption{Energy consumption breakdown of a \textit{typical} mobile operator (data from \cite{GSMA2021_Benchmarking} and \cite{NGMN2021green}).}
\label{fig:EnergyConsumptionBreakdown}
\end{figure}

Given these factors, identifying substantial energy-saving opportunities becomes critical. To enable the required significant savings, an analysis of the exact sources of energy consumption within mobile networks is necessary. As illustrated in Fig.~\ref{fig:EnergyConsumptionBreakdown}, data from \cite{GSMA2021_Benchmarking} and \cite{NGMN2021green} indicates that $73\%$ of the total energy consumption is caused by the \gls{RAN}, predominantly due to \gls{RF} components and associated cooling systems. Notably, cooling alone accounts for approximately $29\%$ of a typical operator's energy consumption, representing more than $0.2\%$ of global electricity usage dedicated solely to cooling base stations \cite{GSMA2019}. Thus, significantly  enhancing energy efficiency must fundamentally influence the \gls{PHY} and the \gls{RF} front end in particular. 

In addition to the need for lower average power consumption, mobile networks face a  traffic challenge. Firstly, data traffic is expected to grow by one to two orders of magnitude in the upcoming ten years \cite{EricssonMobilityReport2024, Gearbox}, which will further exacerbate energy demands unless mitigated by major efficiency improvements. Secondly, traffic loads vary significantly over time and space, with pronounced differences between peak and off-peak hours, as well as between dense urban and rural environments \cite{SpatialTraffic14}. For a network, this leads to a very high probability of low data rates and a very small probability of peak data rates \cite{Gast202403}. This variability, combined with the diversity of use cases, highlights the need for a {flexible PHY} that can adapt energy consumption to instantaneous requirements.\looseness-1

Several existing strategies have been proposed to improve energy efficiency, and each provides valuable contributions at different layers of the network stack. For instance, software-based  AI-driven network management aims to optimize scheduling, load balancing, and traffic steering. These approaches have demonstrated substantial energy savings of  up to $25\%$ in commercial deployments \cite{ericsson2023ai}. However, they cannot directly control the \gls{PHY} and the analog hardware, where a large share of the energy is consumed.

\Gls{ACM}\cite{Goldsmith_ACM}, used in modern communications systems, improves energy efficiency by selecting modulation orders and error correction schemes tailored to channel conditions. While effective in adapting transmission to varying link quality, \gls{ACM} also remains confined to the digital domain and does not address the analog front end.\looseness-1

Another widely discussed strategy is the shutdown of entire base stations during low-traffic periods such as nighttime hours. This method can significantly reduce energy usage but may degrade service availability and short-term agility.

While all of these approaches are beneficial and should be part of a holistic energy efficiency strategy, they do not fully exploit the opportunities that lie in flexible control of the analog front end. The {Gearbox-PHY} approach, introduced in \cite{Gearbox}, is designed to complement these methods by addressing precisely this limitation. Inspired by the analogy of an automatic gearbox in automotive systems, the Gearbox-PHY dynamically selects modulation schemes and corresponding analog front ends based on instantaneous data rate demands and spectral availability. In doing so, it makes better use of the energy-saving potential across the transceiver chain and enables scalable and adaptable operation for future mobile networks.\looseness-1

For example, \gls{IR} modulation can be used to provide a high energy efficiency per communicated bit \cite{Meller202411} for low spectral efficiencies, which is particularly interesting for low data rates or high spectral availability. For required higher data rates under limited spectrum, other schemes, such as \gls{QAM}, become preferable as they can provide higher spectral efficiencies. By switching between schemes and front ends based on requirements the Gearbox-PHY aims to maximize the energy efficiency. 

Due to the anticipated tenfold to hundredfold increase in traffic, \gls{6G} must achieve corresponding tenfold to hundredfold improvements in energy efficiency per bit to maintain the same overall energy consumption as the \gls{5G} of mobile networks at roll-out. This raises a critical question: can a Gearbox-PHY achieve these ambitious targets?  

In this regard, this paper evaluates the energy-saving potential of the Gearbox-PHY. Prior analyses indicate that typical network operation is dominated by low data rates \cite{Gast202403}. We therefore focus on this regime, where increasing energy efficiency is most critical.

To assess the robustness of the Gearbox-PHY beyond idealized assumptions, we incorporate representative hardware impairments. Specifically, we consider finite quantizer resolution and local oscillator phase noise. Both impairments jointly influence front end power consumption and achievable spectral efficiency, thereby enabling the Gearbox-PHY to explicitly trade off communication performance against energy efficiency.
These impairments provide a representative basis to evaluate implementable system configurations without restricting the analysis to ideal hardware assumptions.\looseness-1

Building on this motivation, we formulate the Gearbox-PHY as an energy-per-bit minimization problem that jointly optimizes front end and system parameters, namely bandwidth, duty cycle, and active modulation scheme. The analysis relies on hardware-aware power models for transmitter and receiver, explicitly capturing the trade-offs between quantizer resolution, oscillator phase noise, and power consumption. Phase-noise- and quantizer-resolution-dependent spectral efficiency expressions are derived for \gls{QAM}, \gls{ZXM}, and \gls{IR}, enabling a consistent comparison across fundamentally different modulation schemes. 
To ensure practical relevance, we introduce a multi-stage optimization framework that fixes one analog front end per gear, by selecting a single non-adaptive oscillator, \gls{ADC}, \gls{DAC}, and \gls{LNA}.
Finally, we extend the evaluation from point-to-point links to a cellular setting with spatially distributed users and quantify performance using an area-weighted energy-per-bit metric.\looseness-1

This paper combines previous results \cite{Gast202407,Gast202503} and substantially extends them in scope and depth. Different to \cite{Gast202404}, we consider a broad set of parameters and practical modulation schemes, i.e., \gls{QAM}, \gls{ZXM}, and \gls{IR}, enabling a comprehensive comparison across different gears. Further, we explicitly account for phase noise and fixed \gls{LNA} bandwidth constraints, thereby moving beyond idealized front end assumptions. Moreover, we carry out an initial evaluation of the energy saving potential of the Gearbox-PHY in a cellular setting based on a simplified user distribution, providing first system-level insights that were not addressed previously. Overall, the present work offers a more thorough and practically grounded assessment of the Gearbox-PHY concept.\looseness-1

To assess the energy saving potential of the Gearbox-PHY, Section~\ref{sec:GearboxPHY} first introduces the concept in detail and defines the system model together with the underlying optimization problem. Section~\ref{sec:Gears} then presents the considered gears, comprising the modulation schemes and their associated front end architectures. Power consumption models for the required front end components are presented in Section~\ref{sec:PowerModel}, with the exception of the \gls{LO}, which is analyzed separately in Section~\ref{sec:PhaseNoise} to capture the fundamental trade-off between oscillator power consumption and phase noise characteristics. Subsequently, the influence of the resulting phase noise and quantization on the spectral efficiency is evaluated in Section~\ref{sec:SpectralEfficiency}. Building on these results, the Gearbox-PHY optimization problem is solved using a multi-stage approach in Section~\ref{sec:numresults}, before Section~\ref{sec:Conclusion} concludes the paper.\looseness-1

\section{Gearbox-PHY}
\label{sec:GearboxPHY}
The Gearbox-PHY shifts the \gls{PHY} design objective from maximizing peak data rates to minimizing energy consumption while meeting service requirements. This is achieved by dynamically switching between multiple modulation schemes and their corresponding optimized radio front ends, such that each operating point is served with minimal energy expenditure.
In the original formulation of the concept \cite{Gearbox}, modulation schemes with low spectral efficiency but high energy efficiency, such as \gls{IR}, are identified as favorable in regimes characterized by low data rate demands and abundant spectral resources, while highly spectrally efficient schemes, including high-order \gls{MIMO}-\gls{QAM}, are envisioned for scenarios with limited bandwidth and high required data rates.
Importantly, this adaptive operation does not compromise the achievable data rates of the communication system. Instead, the Gearbox-PHY ensures that the required rates are delivered using the modulation and front end configuration that achieves the highest energy efficiency for the given operating conditions. Thus, the employed modulation schemes need not all be optimized for global energy efficiency.

Evaluating the Gearbox-PHY across multiple carrier frequencies is essential, as the carrier frequency determines available bandwidth, path loss, transmit power requirements, and hardware design constraints. Higher frequencies typically offer wider bandwidths but incur increased path loss and stricter analog front end requirements.
Accordingly, we study four representative carrier frequencies $f_c$: \SI{2.4}{GHz}, \SI{8}{GHz}, \SI{28}{GHz}, and \SI{60}{GHz}, spanning established and emerging bands. This set enables the analysis of diverse bandwidth, propagation, and hardware trade-offs and their impact on the energy optimization of the Gearbox-PHY.

The available spectrum and available transmit time are assumed to be fixed and denoted as $B_\mathrm{max}$ and $T_\mathrm{max}$. However, we allow the system to use less bandwidth $B\leq B_\mathrm{max}$ and less transmit time $T\leq T_\mathrm{max}$ than available, if that is beneficial in terms of energy. 
Using a smaller transmit time than $T_\mathrm{max}$ leads to a \textit{duty cycled} mode of operation, where we denote the duty cycle as $\gamma=\frac{T}{T_\mathrm{max}}\leq 1$. Accordingly, we define an \textit{effective} data rate as $R_\mathrm{eff}=V/T_\mathrm{max}$, with $V$ describing the volume of data to be transmitted in $\SI{}{bit}$. 
For $\gamma<1$ we assume that after the transmission the transmitter enters a sleep mode, while the receiver enters an idle mode, where they consume only a fraction $\epsilon_\mathrm{Tx}$ and $\epsilon_\mathrm{Rx}$ of the power when turned, respectively. In general, we assume $\epsilon_\mathrm{Tx}<\epsilon_\mathrm{Rx}$, as the receiver has to continuously monitor for incoming data and can, thus, not reduce its power as much as the transmitter. 
Next to the constraints on bandwidth $B$ and transmit time $T$, we also limit the average transmit power $P_\mathrm{T}$ to $P_\mathrm{max}$.

Finally, the Gearbox-PHY aims to find the gear and system settings that minimize the energy per bit, which is denoted as
\begin{subequations}
\label{eq:OptimizationProblem}
\begin{alignat}{2}
    \raisetag{10pt}
&\underset{\substack{B, \gamma, P_\mathrm{T}, \text{gear},\\ \text{front end}}}{\text{min}} \!\!\!\!&\mathcal{E}_\mathrm{bit} &\!=\!  \frac{(\gamma+\epsilon_\mathrm{Tx} (1-\gamma))P_\mathrm{Tx}+(\gamma+\epsilon_\mathrm{Rx} (1-\gamma))P_\mathrm{Rx}}{R_\mathrm{eff}} \label{eq:Energy} \\
       & \text{s.t.} & R_\mathrm{eff}&=\gamma\,B\,S \label{eq:dataTransferred}\\
       & & 0  &\leq P_\mathrm{T}\leq P_\mathrm{max}\label{eq:PConstraint}\\
        & & 0  &\leq T\leq T_\mathrm{max}\label{eq:TConstraint}\\
        & & 0 &\leq B\leq B_\mathrm{max} \label{eq:BConstraint},
\end{alignat}
\end{subequations}
where the transmitter power consumption $P_\mathrm{Tx}$, receiver power consumption $P_\mathrm{Rx}$, and the spectral efficiency $S$ all depend on the gear and the corresponding front end hardware in use.

Due to the separation of the transmitter and receiver, a frequency-dependent path loss $L$ is considered that relates the received power $P_\mathrm{R}$ to the transmit power $P_\mathrm{T}$.
Since empirical models such as Okumura-Hata \cite{OkumuraHata} are restricted to limited frequency ranges, and our analysis spans \SI{2.4}{GHz} to \SI{60}{GHz}, we adopt a generic distance-based path loss model, i.e.,
\begin{equation}
P_\mathrm{R}=P_\mathrm{T}/L=P_\mathrm{T} D_\mathrm{R} D_\mathrm{T} \left( \frac{c}{4 \pi f_c d}\right)^\beta, \label{eq:PathLoss}
\end{equation}
where $d$ denotes the link distance, $\beta$ the path loss exponent, $c$ the speed of light, and $D_\mathrm{T}$ and $D_\mathrm{R}$ the transmit and receive antenna gains, respectively.

\section{Gears}
\label{sec:Gears}

\begin{figure*} 
    \centering
      \subfloat[\gls{QAM} Transceiver \label{fig:QAMTransceiver}]{%
       \resizebox{0.23\textwidth}{!}{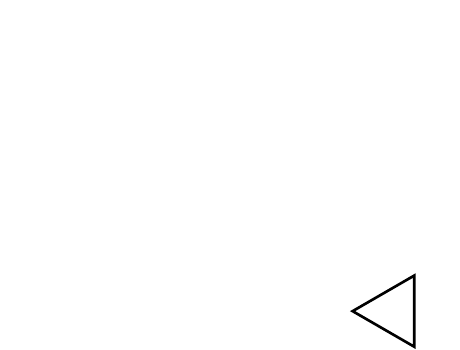} }\hspace{30pt}
      \subfloat[\gls{IR} Transceiver \label{fig:PulseTransceiver}]{%
       \resizebox{0.65\textwidth}{!}{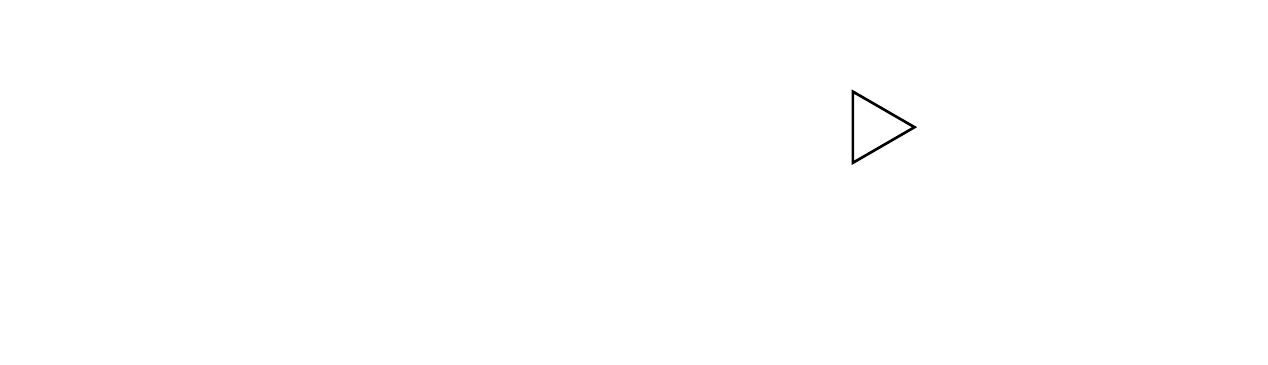}}
       \caption{Assumed Transceiver Architectures \cite{Gast202503}}\vspace{-12pt}
\end{figure*}

We consider three modulation families reflecting distinct design objectives: \gls{QAM} targeting high spectral efficiency, \gls{ZXM} enabling energy-efficient operation at large bandwidths, and \gls{IR} minimizing analog front end complexity.

For all modulations, let $u_k \in \mathcal{U}$ and $w_k \in \mathcal{W}$ denote the transmitted and detected symbols, and let $\mathbf{u} = [u_1,\ldots,u_n]$ and $\mathbf{w} = [w_1,\ldots,w_n]$ denote the corresponding symbol sequences, defining the continuous-time baseband transmit signal as
\begin{equation}
s(t)=\sum_{k} u_k h_\mathrm{Tx}\left(t-\frac{kT_\mathrm{Nyq}}{M_\mathrm{Tx}}\right),
\end{equation}
where $h_\mathrm{Tx}(t)$ denotes the transmit pulse shape, $T_\mathrm{Nyq}$ the Nyquist interval, and $M_\mathrm{Tx}\geq 1$ the \gls{FTN} signaling factor required for \gls{ZXM}. The continuous-time baseband receive signal $r(t)$ is modeled as 
\begin{equation}
r(t)=s(t)e^{j\theta(t)}+\eta(t),\label{eq:signalmodel}
\end{equation}
where $\theta(t)$ represents the phase noise, and $\eta(t)$ is circularly symmetric \gls{AWGN} with variance $\sigma_\eta^2 = N_0 B$. Here $N_0 = k\Gamma$ is the thermal noise \gls{PSD} with Boltzmann constant $k$ and temperature $\Gamma$. 

At the receiver, the signal is filtered by $h_\mathrm{Rx}(t)$, yielding 
$
w(t)=r(t)*h_\mathrm{Rx}(t),
$
and subsequently sampled at rate $M_\mathrm{Rx}/T_\mathrm{Nyq}$ and quantized to obtain $\mathbf{w} = [w_1,\ldots,w_n]$, where $M_\mathrm{Rx} \geq 1$ denotes the receiver oversampling factor. The parameters $M_\mathrm{Tx}$ and $M_\mathrm{Rx}$ control temporal compression and sampling, respectively. Nyquist rate signaling corresponds to $M_\mathrm{Tx}=1$, while $M_\mathrm{Tx}, M_\mathrm{Rx} \geq 1$ are required for \gls{ZXM}.

\subsection{Quadrature Amplitude Modulation}
As an example of linear modulation, we consider the well-known \gls{QAM} family with constellation sizes $M \in \{16, 64, 256, 1024\}$.
The $M$ symbols are uniformly distributed on a rectangular grid. Under hard demapping transmit and receive symbol alphabets $\mathcal{U}$ and $\mathcal{W}$ coincide, defined as
\begin{equation}
    \begin{aligned}
\mathcal{U}\!&=\!\mathcal{W}\!=
\left\{\mathcal{I}\!+\!j\mathcal{Q} | \mathcal{I},\!\mathcal{Q}\!\in \!\left\{\pm 1, \pm 3,..., \pm (\sqrt{M}-1)\right\}\!\right\}\!\!, \label{eq:QAMSymbols}
    \end{aligned}
\end{equation}
with $j$ as the imaginary unit.
We adopt a linear analog transceiver architecture with \glspl{DAC}, an \gls{LO}, a mixer, and a \gls{PA} at the transmitter, and an \gls{LNA}, an \gls{LO}, a mixer, and \glspl{ADC} at the receiver. Passive components, like filters, are assumed to be ideal. The resulting analog front end is shown in Fig.~\ref{fig:QAMTransceiver}.
To minimize quantizer power consumption, we assume hard demapping with a resolution of $\log_2(\sqrt{M})$ bits per I/Q branch for both \glspl{DAC} and \glspl{ADC}. 

\subsection{Zero-Crossing Modulation}
One key challenge of operating at very high bandwidths is the sharp increase in hardware power consumption, particularly of the \gls{ADC}, which scales linearly with the sampling frequency up to roughly $\SI{300}{MHz}$ and quadratically beyond \cite{adcsurvey}, leading to a surge in power for high bandwidths. Nevertheless, ADC power also grows exponentially with amplitude resolution $b_\mathrm{ADC}$ in bits, offering an opportunity for power savings by reducing resolution. However, the performance of conventional modulation schemes, such as high-order QAM, degrades significantly under low-resolution amplitude quantization.\looseness-1

To address this issue, zero-crossing modulation (ZXM) \cite{ZXM} was designed to work with temporally oversampled 1-bit \glspl{ADC}, which compare signals only to a zero threshold, detecting zero-crossings rather than amplitudes. These zero-crossing positions then encode the information in the time domain. This drastically reduces the power consumption and simplifies circuit design, aligning with trends in \gls{CMOS} technologies that favor time-domain resolution over amplitude precision \cite{ziabakhsh2018peak}.

ZXM compensates for the loss in amplitude resolution through \gls{FTN} signaling, where $M_\mathrm{Tx}$ symbols are transmitted per Nyquist interval $T_\mathrm{Nyq}$. To manage the resulting intersymbol interference, run-length-limited (RLL) sequences enforce a minimum distance of $d_\mathrm{RLL}$ symbols between consecutive zero-crossings, with $d_\mathrm{RLL} = M_\mathrm{Tx} - 1$ \cite{neuhaus2021zero}.
Sampling at the receiver is performed every $T_\mathrm{Nyq}/M_\mathrm{Rx}$, where $M_\mathrm{Rx}$ describes a receiver oversampling factor, leading to an effective oversampling of $M_\mathrm{Rx}/M_\mathrm{Tx} \geq 1$. Here, we assume $M_\mathrm{Rx}=M_\mathrm{Tx}$ to avoid equalization and additional digital front end complexity.

The ZXM front end is largely equivalent to the QAM architecture in Fig.~\ref{fig:QAMTransceiver}, except for 1-bit \glspl{DAC} and \glspl{ADC} operating in the I/Q branches. When $M_\mathrm{Tx} > 1$, these components operate at higher than Nyquist rate. Note that, if $M_\mathrm{Tx}= M_\mathrm{Rx} = 1$, ZXM is equivalent to 4-QAM with hard demapping.\looseness-1
\subsection{Impulse Radio}
At low data rates, large distances between zero-crossings in \gls{ZXM} lead to prolonged signal holds, which are energy-inefficient. Since information is encoded in time, this motivates to concentrate energy close to the timings of the zero crossings, which leads to \gls{IR}.
We focus on real-valued IR to minimize front end complexity, prioritizing energy efficiency over capacity. Here, we evaluate two schemes with differing use of phase information: variable-sign IR and unipolar IR \cite{Roth202409}.\looseness-1

{Unipolar IR} transmits pulses of only a single polarity, fully discarding phase encoding and relying solely on timing. This simplifies the receiver architecture by removing the need for downconversion, i.e., no local oscillator or mixer is needed. The front end for unipolar IR (Fig.~\ref{fig:PulseTransceiver}) includes a 1-bit \gls{DAC}, a single-ended \gls{LO}, a switch, a filter, and a \gls{PA} at the transmitter. The receiver comprises an \gls{LNA}, a filter, a power detector, and an \gls{ADC}. Operating directly on the high-frequency signal, the power detector makes the system inherently robust to phase rotations and further reduces power consumption.

In {variable-sign IR}, pulse amplitudes may take positive or negative values, encoding information both in timing and phase. This approach achieves higher spectral efficiency across all considered \glspl{SNR} compared to unipolar IR. The transmitter consists of a 3-level \gls{DAC}, a single-ended \gls{LO}, a 3-way switch (positive, negative, or no pulse), a filter, and a \gls{PA}. The receiver includes an \gls{LNA}, a single-ended \gls{LO}, a mixer, a passive filter, and a 3-level \gls{ADC}, corresponding to approximately $\SI{1.59}{bit}$ resolution (Fig.~\ref{fig:PulseTransceiver}).\looseness-1

\section{Hardware Power Consumption Model}
\label{sec:PowerModel}

To quantify the hardware power consumption as a function of the system parameters in (\ref{eq:OptimizationProblem}), we consider the active components of the analog front end. Each block is modeled individually based on circuit-level insights or empirical data. In the following, we detail the power models used for each of these components, which is similar to \cite{Gast202404}, with refinements for the \gls{ADC}, \gls{DAC}, \gls{PA}, \gls{LNA}, and \gls{LO}.
\subsection{Digital-to-Analog Converter}
We model the \gls{DAC} based on a binary current-steering architecture, which is used for high-speed, high-resolution applications \cite{DACSurvey}. Assuming Nyquist-rate sampling, the power consumption for such an architecture is given in \cite{CuiOptimization}.\looseness-1

However, oversampling at the \gls{DAC}, to realize FTN signaling via $M_\mathrm{TX}>1$ for \gls{ZXM}, increases its power consumption, though it does not expand the required bandwidth. Following \cite{neuhaus2021EE}, we assume a time-interleaved \gls{DAC} architecture, where multiple converters operate in parallel. 
This motivates the power consumption model from \cite{CuiOptimization} to be extended as
\begin{align}
    P_\mathrm{DAC}& >  M_\mathrm{Tx} \left( \frac{1}{2} V_\mathrm{dd} I_\mathrm{0}  (2^b_\mathrm{DAC} - 1)+ C_\mathrm{p} V_\mathrm{dd}^2 b_\mathrm{DAC}  B \right),
\end{align} 
where $V_\mathrm{dd}$ is the supply voltage, $I_0$ the unit current source, $C_\mathrm{p}$ the parasitic capacitance, and $b_\mathrm{DAC}$ the resolution in bits. 
\subsection{Mixer}
Mixers are required at both the transmitter and receiver for frequency up- and down-conversion. We consider complex implementations and note that no general expression exists for mixer power consumption. Also w.r.t. design parameters like bandwidth $B$, surveys of CMOS implementations show no consistent trend \cite{zhang2022optimal}. In practice, mixer power is highly implementation-dependent. We therefore adopt measurement-based values from the literature, as summarized in Table~\ref{tab:MixerLOPow}.
\begin{table}
\center
\caption{Literature-based measurements for mixer power consumption}\vspace{-6pt}
\resizebox{\columnwidth}{!}{%
\begin{tabular}{|l|l|l|l|l|}
\hline
$f_c$                    & $\SI{2.4}{GHz}$ &  $\SI{8}{GHz}$ & $\SI{28}{GHz}$ & $\SI{60}{GHz}$  \\ \hline
$P_\mathrm{Mix}$                    &  $\SI{1.57}{mW}$ \cite{MuradMixer2GHz} & $\SI{6.9}{mW}$ \cite{mahmoudi20048} & $\SI{8.4}{mW}$ \cite{chang202128}   &  $\SI{17}{mW}$ \cite{lee201360} \\\hline
\end{tabular}
}
\label{tab:MixerLOPow}
\end{table}
\subsection{Power Amplifier}
\gls{PA} power consumption is highly dependent on the class, implementation, and technology, such that no general model for the power consumption is available to the best of our knowledge. Instead, we adopt an empirical lower-bound based on the \gls{PAE}, a standard \gls{PA} metric, defined as \cite[Chapter 2]{wang2015rf}
\begin{equation}
P\!AE = \frac{P_\mathrm{out} - P_\mathrm{in}}{P_\mathrm{PA}}, \label{eq:PAE}
\end{equation}
where $P_\mathrm{in}$ and $P_\mathrm{out}$ denote the input and output power of the PA, and $P_\mathrm{PA}$ its total power consumption. Since in most cases $P_\mathrm{out} \gg P_\mathrm{in}$, we approximate $P_\mathrm{PA} \approx P_\mathrm{out}/P\!AE$. Assuming ideal antennas and matching and, thus, $P_\mathrm{out} = P_\mathrm{T}$, this yields
\begin{equation}
P_\mathrm{PA} \approx \frac{P_\mathrm{T}}{P\!AE}. \label{eq:P_PA_approx}
\end{equation}
To model the PAE, we reference the survey \cite{PASurvey}, which reports maximum achievable PAE values, $P\!AE_\mathrm{max}$, for various \glspl{PA} at different carrier frequencies $f_c$. Following \cite{Gast202404}, fitting the data from \cite{PASurvey} with an L1-regularized regression yields
\begin{equation}
P\!AE_\mathrm{max} \approx 0.732 \left( \frac{f_c}{\SI{1}{GHz}} \right)^{-0.5}. \label{eq:PAE_fit}
\end{equation}
Substituting (\ref{eq:PAE_fit}) into (\ref{eq:P_PA_approx}) provides a lower bound for the PA power consumption as
\begin{equation}
P_\mathrm{PA} \gtrapprox 4.32 \cdot 10^{-5} \,\mathrm{Hz}^{-0.5} \cdot P_\mathrm{T} \cdot \sqrt{f_c}, \label{eq:P_PA_lower}
\end{equation}
which corresponds to operation at the maximum efficiency point of the \gls{PA} and, thus, constitutes an optimistic lower bound on $P_\mathrm{PA}$. 

In practice, the achievable \gls{PAE} is further reduced due to linearity constraints. Most modulation schemes require operation with output backoff from saturation, which is governed by the \gls{PAPR} $\Psi$ and lowers the average \gls{PAE} \cite[Chapter~2]{wang2015rf}. To capture this effect while retaining analytical tractability, we model the effective \gls{PAE} as a fraction of the maximum achievable value and approximate the degradation as inversely proportional to $\Psi$, i.e.,
\begin{equation}
P\!AE \gtrapprox \frac{P\!AE_\mathrm{max}}{\Psi}. \label{PAE_PAPR_Relation}
\end{equation}
This yields a conservative approximation of the \gls{PAE}.
Note that this backoff assumption primarily applies to linear modulation schemes, where distortion must be avoided. In contrast, schemes like \gls{ZXM} tolerate significant non-linearities, see e.g. \cite{Gast202503b}, enabling operation closer to saturation. This is especially advantageous at high frequencies, where \glspl{PA}  have severely limited output power \cite{PASurvey}, which is neglected here.
Including the \gls{PAPR} $\Psi$ into (\ref{eq:P_PA_lower}), we model the \gls{PA} power consumption as\looseness-1
 \begin{equation}
     P_\mathrm{PA} \approx 4.32 \cdot 10^{-5} \mathrm{Hz}^{-0.5} P_\mathrm{T} \sqrt{f_c} \Psi, \label{eq:PApowerWithPAPR}
 \end{equation}

To evaluate (\ref{eq:PApowerWithPAPR}), we require the \gls{PAPR} $\Psi$ of each modulation scheme, which can be analyzed using the \gls{PMEPR}, a baseband-equivalent metric that simplifies the analysis. For complex baseband signals and sufficiently high carrier frequencies $f_c$, the \gls{PAPR} relates to the \gls{PMEPR} as \cite{PAPR}
\begin{equation}
\left.\Psi(s(t))\right|_\mathrm{dB} = \left.PMEPR(s(t))\right|_\mathrm{dB} + \SI{3}{dB}, \label{eq:PMEPR2PAPR}
\end{equation}
where $s(t)$ is the continuous-time baseband transmit signal.

\paragraph*{PAPR of \gls{QAM}}
For linear modulation with a fixed roll-off $\alpha$, the \gls{PMEPR} equals the symbol-\gls{PMEPR} $\Upsilon$, defined as the peak-to-average symbol energy in the I/Q plane with an offset introduced by filtering \cite{PAPR}. For a \gls{RRC} filter with $\alpha = 0.5$, this offset is approximately \SI{3.17}{dB} \cite{PAPR}, while the symbol-\gls{PMEPR} for $M$-\gls{QAM} is given as \cite{wolf2010maximum}\looseness-1
\begin{equation}
\Upsilon_\mathrm{M\text{-}QAM} = \frac{3(\sqrt{M} - 1)}{\sqrt{M} + 1}.
\end{equation}

\paragraph*{PAPR of \gls{ZXM}}
The PMEPR of \gls{ZXM} cannot be evaluated symbol-wise due to FTN signaling. However, as shown in \cite{neuhaus2021zero}, using the same \gls{RRC} filtering with $\alpha = 0.5$,  it remains constant at roughly \SI{3.63}{dB}, independent of the FTN factor $M_\mathrm{Tx}$.

\paragraph*{PAPR of \gls{IR}}
Following  the discussion in \cite[Ch. 6]{6GlifeBook}, we replace \gls{RRC} filters with raised-cosine (RC) filters for \gls{IR} to improve spectral efficiency, using a roll-off of $\alpha = 0.25$. As RC filters at the transmitter and the receiver do not meet the Nyquist {ISI} criterion, symbol-wise PAPR evaluation is infeasible and we determine the \gls{PAPR} values numerically.\looseness-1

The resulting \glspl{PAPR} $\Psi$ for all considered schemes are summarized in Table~\ref{tab:PAPR}. These values are used in (\ref{eq:PApowerWithPAPR}) to estimate the corresponding PA power consumption and assess the impact of the PAPR on system energy efficiency.

    \renewcommand{\arraystretch}{1.8}
\begin{table}
    \centering
    \caption{PAPRs $\Psi$ of considered modulation schemes}
    \vspace{-6pt}
    \begin{tabular}{|c|c|}
    \hline 
    Modulation scheme   & PAPR $\Psi$ \\ \hline  
    $M$-QAM     &  $10 \log_{10}\frac{3(\sqrt{M}-1)}{\sqrt{M}+1}\,\mathrm{dB}+\SI{3.17}{dB}+\SI{3}{dB}$\\ \hline
       \gls{ZXM}  & $\SI{3.63}{dB}+\SI{3}{dB}$ \\ \hline
       unipolar \gls{IR} & $\SI{6.48}{dB}$ \\ \hline 
       variable-sign \gls{IR} & $\SI{7.72}{dB}$ \\ \hline 
    \end{tabular}
    \label{tab:PAPR} 
\end{table}

\subsection{Low-Noise Amplifier}
The power consumption of an \gls{LNA} depends on design goals such as noise minimization or dynamic range, typically captured by various \glspl{FOM} \cite{LNAFoMs}. For system-level analysis, we adopt a bandwidth-dependent \gls{FOM}, assuming high-gain, low-distortion operation so that power scales primarily with bandwidth.
Following \cite{Mezghani_PowerEfficiency}, the LNA power is, thus, modeled via a \gls{FOM} as
\begin{equation}
FOM_\mathrm{LNA} = \frac{G_\mathrm{LNA} B N_0}{(N_\mathrm{LNA} - 1) P_\mathrm{LNA}},
\end{equation}
where $G_\mathrm{LNA}$ is the gain, $N_\mathrm{LNA}$ the noise figure, $N_0$ the noise PSD, and $P_\mathrm{LNA}$ the power consumption.

This \gls{FOM} is considered technology-independent with reported values ranging from $10^{-7}$ to $10^{-9}$ \cite{Mezghani_PowerEfficiency}. In the following, we adopt a fixed noise figure of \SI{5}{dB} and a gain between \SI{10}{dB} and \SI{20}{dB}, consistent with \cite{LNASurveyPaper}.

\subsection{Analog-to-Digital Converter}
Although \glspl{ADC} and \glspl{DAC} both perform signal-domain conversions, their power characteristics differ greatly. At low bandwidths, \glspl{DAC} tend to dominate power consumption, while at high bandwidths, \glspl{ADC} become more power-intensive \cite{bossy2020flexible}.
We model the ADC power consumption based on the Walden \gls{FOM}, defined in  \cite{ADCSurveyKeynotePaper} as
\begin{equation}
FOM_\mathrm{W} = \frac{P_\mathrm{ADC}}{f_s 2^{ENOB}}, \label{eq:FOM_def}
\end{equation}
with $f_s$ being the Nyquist sampling rate and $ENOB$ the effective number of bits. We approximate $ENOB \approx b_\mathrm{ADC}$ and assume Nyquist-rate sampling, i.e., $f_s = B$. A survey of state-of-the-art ADCs \cite{adcsurvey} provides an empirical lower bound as\looseness-1
\begin{equation}
FOM_\mathrm{W} \geq 0.67 \cdot 10^{-15} \sqrt{1 + \left(\frac{B}{\SI{560}{MHz}}\right)^2} \mathrm{J/conv\text{-}step}, \label{eq:MurmannEnvelope}
\end{equation}
yielding a pessimistic lower bound on ADC power as
\begin{equation}
P_\mathrm{ADC} \gtrapprox 0.67 \cdot 10^{-15} \cdot 2^{b_\mathrm{ADC}} B \sqrt{1 + \left( \frac{B}{\SI{560}{MHz}} \right)^2}. \label{eq:P_ADC}
\end{equation}

For \gls{ZXM}, temporal oversampling is handled via time-interleaved architectures, commonly used in oscilloscopes and suitable for time-domain signaling \cite{neuhaus2021EE}. The oversampling factor $M_\mathrm{Rx}$ scales the number of parallel converters but does not increase bandwidth. Thus, the lower-bound power model in (\ref{eq:P_ADC}) is extended as
\begin{align}
P_\mathrm{ADC} &\!> \!M_\mathrm{Rx} \!\cdot\! 0.67 \!\cdot\! 10^{-15} \!\cdot \! 2^{b_\mathrm{ADC}} B \sqrt{1 \!+\!\! \left(\! \frac{B}{\SI{560}{MHz}} \!\right)\!^2}\!, \label{eq:P_ADC_MTX} 
\end{align}
where signal recombination overheads are not included.
\subsection{Power Detector}
The power detector in Fig.~\ref{fig:PulseTransceiver} plays a central role in the overall energy consumption of unipolar \gls{IR} receivers. Usually, this component can be realized with high efficiency. Multiple implementations have demonstrated exceptionally low power requirements. For example, the \SI{60}{GHz} detector in \cite{PowerDetector} achieves operation at only \SI{1.6}{mA} from a \SI{1.5}{V} supply, corresponding to \SI{2.4}{mW} of power consumption. Based on this figure, we conservatively model the power consumption of the power detector as $P_\mathrm{PD} = \SI{2.4}{mW}$ across all frequencies. The detector acts as a \gls{RSSI} unit, providing an estimate of the received signal strength.
\subsection{Transmitter and Receiver Power}
\label{sec:RealModulationSchemes:Transmitter_Receiver_Power}

The transmitter and receiver power consumptions, denoted as $P_\mathrm{Tx}$ and $P_\mathrm{Rx}$, are calculated using the models introduced in this section.
Specifically, $P_\mathrm{Tx}$ is calculated as
\begin{equation*}
	P_\mathrm{Tx}\!=\!
	\begin{cases} 
	2P_\mathrm{DAC}\!+\!P_\mathrm{LO}\!+\!P_\mathrm{Mix}\!+\!P_\mathrm{PA} & \text{for \gls{QAM},  \gls{ZXM}}\\
	P_\mathrm{DAC}\!+\!P_\mathrm{LO}\!+\!P_\mathrm{Mix}\!+\!P_\mathrm{PA} & \text{for \gls{IR}},
	\end{cases}
\end{equation*}
while the receiver power $P_\mathrm{Rx}$ is modeled as
\begin{equation*}
	P_\mathrm{Rx}\!=\!
	\begin{cases} 
	P_\mathrm{LNA}\!+\!P_\mathrm{LO}\!+\!P_\mathrm{Mix}\!+\!2P_\mathrm{ADC} & \!\!\text{for \gls{QAM},  \gls{ZXM}}\\
	P_\mathrm{LNA}\!+\!P_\mathrm{LO}\!+\!P_\mathrm{Mix}\!+\!P_\mathrm{ADC} & \!\!\text{for variable-sign \gls{IR}}\\
	P_\mathrm{LNA}\!+\!P_\mathrm{PD}\!+\!P_\mathrm{ADC} & \!\!\text{for unipolar \gls{IR}},
	\end{cases}\hspace{-10pt}
\end{equation*}
where $P_\mathrm{LO}$ is the power consumption of the \gls{LO}, which will be analyzed in depth in the next chapter.
\section{Phase Noise - Power Trade-Off}
\label{sec:PhaseNoise}

As explained in the introduction, we incorporate phase noise as a key hardware impairment in our optimization. As such, we dedicate a section solely to modeling the phase noise and the power of the local oscillator.
\subsection{Phase Noise Model}
\label{sec:pnmodel}
We start by introducing a phase noise model, before connecting it with the power consumption of the \gls{LO}.
For this, we adopt the model in \cite{KhanzadiPhaseNoiseModel}, which provides a tractable phase noise model aligned with oscillator measurements. Similar to other models, it is based on the single-sideband phase noise spectrum $\mathcal{L}(f_m)$, defined as the power within a $\SI{1}{Hz}$ bandwidth at an offset $f_m$ from the carrier $f_c$, normalized by the total output power. For $f_m$ beyond the oscillator line-width, i.e., the $3$-dB corner of $\mathcal{L}(f_m)$, $\mathcal{L}(f_m)$ approximates the phase noise PSD $S_\theta(f_m)$ \cite{KhanzadiPhaseNoiseModel}.

In \cite{KhanzadiPhaseNoiseModel}, the phase noise $\theta (t)$ is modeled as a sum of three independent processes
\begin{equation}
\theta(t)=\theta_3(t) + \theta_2(t)+\theta_0(t),
\end{equation}
with spectra 
$K_3/(f_m^3+f_\mathrm{PLL}^3)$, $K_2/(f_m^2+f_\mathrm{PLL}^2)$ for $\theta_3(t)$ and $\theta_2(t)$, respectively, and flat noise $K_0$ for $\theta_0(t)$, where $f_\mathrm{PLL}$ describes the \gls{PLL} bandwidth. These terms originate from flicker noise, integrated white noise, and thermal noise, respectively. The component $\theta_2(t)$, with a $1/f_m^2$ slope, is a widely considered phase noise representation and corresponds to a Wiener process \cite{KhanzadiPhaseNoiseModel}.

The model considers \gls{PLL} stabilization, where a free-running \gls{VCO} is locked to a low-frequency reference oscillator. A typical \gls{PLL} architecture includes a frequency divider, a phase-frequency detector, a charge pump, and a loop filter to control the \gls{VCO}. Although all components contribute to phase noise \cite{JitterPower}, the \gls{VCO} dominates at higher $f_m$, and is, thus, the primary focus in \cite{KhanzadiPhaseNoiseModel}. The PLL is approximated as a high-pass filter with cutoff frequency $f_\mathrm{PLL}$, attenuating phase noise below this threshold and resulting in a flattened PSD for $f_m<f_\mathrm{PLL}$. 
Consequently, if $f_\mathrm{PLL}$ is sufficiently large, considering the PSD of the oscillator, $\theta_3(t)$ can be typically neglected. In our analysis, we assume $f_\mathrm{PLL}=\SI{1}{MHz}$
\cite{Szortyka2024PLL}.\looseness-1

\subsection{\gls{LO} Power Consumption}

To model the power consumption of the local oscillator, we follow the approach in \cite{JitterPower}, which derives a power model for the \gls{VCO} in an integer-$N$ PLL. In this model, the phase noise spectrum at the VCO output is assumed to be flat up to the loop bandwidth $f_\mathrm{PLL}$, followed by a $1/f_m^2$ decay, resulting in a finite total phase noise power even over infinite bandwidths and, thus, neglects $\theta_3$ and $\theta_0$.

This spectral shape corresponds to a Wiener-type phase noise process with a \gls{PSD} of the form $\frac{K_2}{f_\mathrm{PLL}^2 + f_m^2}$, leading to a total phase noise power
\begin{equation}
\sigma^2_J = \int_{-\infty}^{\infty} \frac{K_2}{f_\mathrm{PLL}^2 + f_m^2} \mathrm{d}f_m = \frac{K_2 \pi}{f_\mathrm{PLL}}. \label{eq:WienerPower}
\end{equation}

Note that this result differs from \cite[eq.~(3)]{JitterPower} due to two factors: (i) a closed-form integral is used instead of an approximation, and (ii) we express phase noise in radians, while \cite{JitterPower} uses jitter in seconds. These two representations are equivalent and related by the carrier frequency $f_c$ as
\begin{equation}
\sigma_J\mathrm{[rad]} = \sigma_J\mathrm{[s]} \cdot 2\pi f_c. \label{eq:ConversionPhaseNoiseJitter}
\end{equation}

The \gls{VCO} power consumption in \cite[eq.~(26)]{JitterPower} can be expressed as
\begin{equation}
P_\mathrm{VCO} = \frac{k \Gamma (1+\delta)}{\pi^2 Q_\mathrm{O}^2 f_\mathrm{REF}^2} \left( S_\mathrm{REF} + S_\mathrm{CP} \right) \left( \frac{2\pi f_c}{\sigma_J\mathrm{[rad]}} \right)^4, \label{eq:VCO_Power_Jitter}
\end{equation}
where $f_\mathrm{REF}$ is the reference oscillator frequency, $k$ is the Boltzmann constant, $\Gamma = \SI{300}{K}$ is the operating temperature, $\delta = 1$ accounts for excess noise in MOS devices, and $Q_\mathrm{O} = 10$ is the assumed oscillator quality factor \cite{JitterPower}. The terms $S_\mathrm{REF}$ and $S_\mathrm{CP}$ denote the white-noise \glspl{PSD} of the reference oscillator and charge pump, respectively. Following \cite[Sec. VI]{JitterPower}, we set $S_\mathrm{CP} = S_\mathrm{REF}$. Equation \eqref{eq:VCO_Power_Jitter} highlights that the \gls{LO} power scales inversely with the fourth power of the phase noise standard deviation, i.e., $P_\mathrm{LO} \sim 1/\sigma_J^4$, underscoring the high cost of achieving low phase noise.

Here, we explicitly include the reference oscillator power consumption. Based on the crystal oscillator in \cite{Crystal}, we set $P_\mathrm{REF} = \SI{198}{\mu W}$ with a reference frequency of $f_\mathrm{REF} = \SI{54}{MHz}$ and an approximate flat \gls{PSD} of $S_\mathrm{REF} = \SI{-160}{dBc/Hz}$ for offset frequencies $f_m > \SI{10}{kHz}$.

Combining the dominant contributors, we define the total power consumption of the local oscillator as

\begin{equation}
P_\mathrm{LO} \geq P_\mathrm{VCO} + P_\mathrm{REF} \label{eq:LOFullPowerModel}
\end{equation}
neglecting the relatively small contributions from the phase-frequency detector, charge pump, and frequency divider.

\subsection{Resulting Phase Noise}
We now connect the \gls{VCO} power model from \cite{JitterPower} in (\ref{eq:LOFullPowerModel}) with the phase noise model from \cite{KhanzadiPhaseNoiseModel}. Since \cite{JitterPower} derives the VCO power consumption solely based on the Wiener phase noise component $\theta_2(t)$, we omit the cubic term $\theta_3(t)$ for consistency. This is a common simplification, particularly valid for sufficiently large PLL loop bandwidths $f_\mathrm{PLL}$, where the impact of $\theta_3(t)$ is effectively suppressed.
The model in \cite{JitterPower} further does not address the oscillator noise floor $K_0$. However, as this term dominates for wideband systems, we explicitly include it. Reported values in the literature range from $\SI{-140}{dBc/Hz}$ to $\SI{-125}{dBc/Hz}$ even for sub-THz designs, see e.g.,  \cite{li2018200,laemmle2013fully}. To remain conservative, we adopt $K_0 = \SI{-125}{dBc/Hz}$ in the following.

Given a Wiener phase noise power $\sigma_J^2$, the corresponding \gls{VCO} power can be determined via \eqref{eq:VCO_Power_Jitter}. By using \eqref{eq:WienerPower} to obtain $K_2$, we express the full phase noise \gls{PSD} as

\begin{equation}
\label{eq:PhaseNoisePSD}
S_\theta(f_m) = 2\frac{\sigma_J^2 f_\mathrm{PLL} / \pi}{f_\mathrm{PLL}^2 + f_m^2} + 2K_0,
\end{equation}
where the factor 2 accounts for the use of identical oscillators at both transmitter and receiver. 

\subsection{Phase Noise Tracking}
Having linked the Wiener phase noise variance $\sigma_J^2$ to the \gls{LO} power consumption in (\ref{eq:VCO_Power_Jitter}) and to the complete phase noise \gls{PSD} in (\ref{eq:PhaseNoisePSD}), we now consider phase noise tracking and correction at the receiver as a means to mitigate phase noise effects for a given $\sigma_J^2$.

Tracking capabilities differ by system architecture. While high resolution \glspl{ADC} allow for the application of classical Wiener interpolation, as detailed in \cite{SimonPhaseNoise}, receivers with 1-bit quantization, such as those for \gls{ZXM}, pose greater challenges for phase tracking. 
However, results like \cite{Gast202307} confirm that phase tracking remains viable even at 1-bit resolution. 
As discussed, the two considered \gls{IR} schemes employ real-valued modulation. As such, for variable-sign \gls{IR}, phase noise reduces signal amplitude, while unipolar \gls{IR} is unaffected due to power detection. Hence, phase tracking in the usual sense is considered to be not applicable to these schemes.

The phase $\theta(t)$ affects the signal multiplicatively via the phasor $e^{j\theta(t)}$ in (\ref{eq:signalmodel}). While $\theta(t)$ is non-stationary, the phasor process is wide-sense stationary. Assuming $\theta(t)\ll 1$, the \gls{PSD} of $\theta(t)$  approximates that of the phasor $e^{j\theta(t)}$ \cite{meyr1998digital}.

We consider pilot-based tracking with every $F$-th symbol as a known pilot, reducing the effective data rate to $\frac{F-1}{F} R_\mathrm{eff}$. For bandwidth $B$, we consider a pilot spacing of roughly $\frac{F}{B}$, enabling reconstruction of frequency components below $\frac{B}{2F}$. Thus, while more pilots lower throughput, they improve phase estimation accuracy.
To quantify this trade-off, we decompose $\theta(t) = \theta_\mathrm{L}(t) + \theta_\mathrm{H}(t)$, where $\theta_\mathrm{L}(t)$ contains low-frequency, trackable components with $f_m < \frac{B}{2F}$, and $\theta_\mathrm{H}(t)$ contains the untrackable high-frequency portion. Assuming \gls{LMMSE} interpolation, the \gls{PSD} of the residual phase estimation error is given by \cite{doerpinghaus2012Joint} as\looseness-1
\begin{equation}
\label{eq:ErrorSpectrum}
S_e(f_m)=\frac{S_{\theta_L}(f_m)}{\frac{\xi}{F} B S_{\theta_L}(f_m)+1}+S_{\theta_H}(f_m),
\end{equation}
with $\xi=\frac{\sigma_s^2}{\sigma_\eta^2}$, where $\sigma_s^2$ and $\sigma_\eta^2$ denote the power of the transmit signal $s(t)$ and the noise $\eta(t)$, respectively. 
Since (\ref{eq:ErrorSpectrum}) neglects quantization \cite{doerpinghaus2012Joint}, we model its effect as an \gls{SNR} degradation, i.e., we evaluate $\xi^\prime=\frac{\sigma_s^2}{\sigma_\eta^2+\sigma_q^2}$, where $\sigma_q^2$ is the quantization noise power. Following \cite{DitheredQuant}, we set
$
\frac{\sigma_q^2}{2}=\frac{\Delta^2}{4},
$
with the step size $\Delta$ given as $\Delta=\frac{2 \Xi}{2^b_\mathrm{ADC}}$, where $b_\mathrm{ADC}$ denotes the \gls{ADC} resolution and  $\Xi$ describes the one-sided dynamic range. We express $\Xi$ as a multiple $\kappa$ of the \gls{ADC} input variance, i.e., $\Xi=\sqrt{\kappa(\sigma_s^2 + \sigma_n^2)/2}$, which allows us to denote the degraded SNR as
\begin{equation}
\label{eq:EffectiveSNR}
\xi'=\left(\frac{\sigma_\eta^2}{\sigma_s^2}+\frac{\kappa\left(1+\frac{\sigma_\eta^2}{\sigma_s^2}\right)}{2^{b_\mathrm{ADC}}} \right)^{-1}.
\end{equation}
In the following we fix $\kappa=2$, corresponding to an overload probability below $5\%$.

\begin{figure}
    \centering
    \resizebox{0.95\columnwidth}{!}{\import{figures}{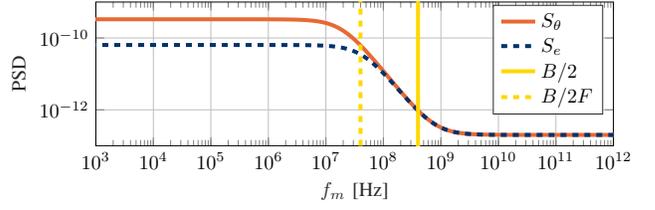}} 
    \caption{Exemplary \glspl{PSD} for $\sigma_J^2=0.1$, $K_0=\SI{-125}{dBc/Hz}$, ADC bits $b_\mathrm{ADC}=10$, $\kappa=2$, bandwidth $B=\SI{400}{MHz}$, pilot spacing $F=10$ \cite{Gast202503}}
    \label{fig:ExamplePSD}
\end{figure}

With $\xi'$ from (\ref{eq:EffectiveSNR}) and $S_\theta(f_m)$ from (\ref{eq:PhaseNoisePSD}) inserted into (\ref{eq:ErrorSpectrum}), we compute the corrected phase noise spectrum after tracking. An example is shown in Fig.~\ref{fig:ExamplePSD}, which showcases the mitigation of low-frequency components.
The residual phase noise variance $\sigma_\mathrm{PN}^2$ after tracking can then be calculated as the integral of (\ref{eq:ErrorSpectrum}) over the bandwidth $B$, i.e.,
\begin{align}
\label{eq:resultingPNVar}
\sigma_\mathrm{PN}^2=&2\int_0^{\frac{B}{2F}}\!\! \frac{S_\theta(f_m)}{\frac{\xi' B}{F} S_\theta(f_m)+1} \mathrm{d}f_m\! + \!2\int_{\frac{B}{2F}}^{B/2} \!\!S_\theta(f_m)\mathrm{d}f_m . \raisetag{6pt}
\end{align}
It is not intuitively clear whether (\ref{eq:resultingPNVar}) holds for the extreme case of 1-bit quantization. However, as shown in \cite{Zeit202212}, estimators using 1-bit samples \cite{Gast202205,Gast202307} can achieve the Bayesian Cramér-Rao lower bound, though based on a block-based model with pilot and data segments. Still, these results support the assumption that (\ref{eq:ErrorSpectrum}) and (\ref{eq:resultingPNVar}) remain valid even under extreme quantization.
For the real-valued \gls{IR} schemes, where no tracking is considered, the phase noise variance $\sigma^2_\mathrm{PN}$ in (\ref{eq:resultingPNVar}) can be calculated by setting $F\to \infty$.

\section{Spectral Efficiency}
\label{sec:SpectralEfficiency}
In order to evaluate the energy per bit for different modulation schemes, front end configurations, and hardware impairments, we analyze the respective influences on the spectral efficiency.  
We define the spectral efficiency based on the normalized mutual information rate between the transmit symbols $\mathsf{U}$ and the received symbols $\mathsf{W}$ normalized by the $99\%$ containment bandwidth $B_{99}$, i.e.,
\begin{equation}
S = I^\prime(\mathsf{U};\mathsf{W})/B_{99}.
\end{equation}
Assessing the impact of correlated phase noise on the spectral efficiency poses significant analytical challenges. To obtain a tractable lower bound on the achievable spectral efficiency, we evaluate the worst-case scenario by modeling the residual phase noise error as white Gaussian noise, i.e.,  $\theta_e[k] \sim \mathcal{N}(0,\sigma_\mathrm{PN}^2)$, with variance $\sigma_\mathrm{PN}^2$ defined in (\ref{eq:resultingPNVar}), leading to a lower bound on the spectral efficiency expressions.
\subsection{Information Rate of \gls{QAM}}
We assume the \gls{DAC}  and the  \gls{ADC} to have equal minimal amplitude resolutions, i.e., setting $b_\mathrm{DAC}=b_\mathrm{ADC}=\frac{1}{2}\log_2(M)$ per I/Q dimension, which corresponds to hard demapping.
As we can evaluate the mutual information rate on a per symbol basis, due to the memoryless structure, it can be expressed as
\begin{equation}    
I^\prime(\mathsf{U};\mathsf{W})\!=\!\sum_{u \in \mathcal{U}} \sum_{w \in \mathcal{W}} P_{\mathsf{U},\mathsf{W}}(u,w) \log_2\! \left( \frac{P_{\mathsf{W}|\mathsf{U}}(w|u)}{\sum_{u \in \mathcal{U}} \frac{1}{M}P_{\mathsf{W}|\mathsf{U}}(w|u)}\right)\!, \label{eq:MUIforQAM}
\end{equation}
assuming a uniform distribution of the transmit symbols. 
Here, the probabilities $P_{\mathsf{W}|\mathsf{U}}$ include the effects of the residual phase noise error, modeled as $\theta_e[k]\sim \mathcal{N}(0, \sigma_\mathrm{PN}^2)$ as per (\ref{eq:resultingPNVar}). 
Since a closed-form evaluation of $P_{\mathsf{W}|\mathsf{U}}$ is intractable, it is approximated by averaging over $N$ phase-noise realizations, i.e.,\looseness-1
\begin{equation}
P_{\mathsf{W}|\mathsf{U}}(w|u)=\frac{1}{N}\sum_k P_{\mathsf{W}|\mathsf{U},\theta}(w|u e^{j\theta_e[k]}). \label{eq:proboverpn}
\end{equation}

The additive noise is modeled as circularly symmetric complex Gaussian with variance $\sigma_\eta^2$. Hard decision demapping is assumed and decision regions are defined by upper and lower thresholds corresponding to the midpoints between adjacent constellation points in $\mathcal{I}$ and $\mathcal{Q}$ in~\eqref{eq:QAMSymbols}, with the outermost regions extending to $\pm\infty$. For the real component of symbol $w \in \mathcal{W}$, these thresholds are denoted by $d_{\Re w,u}$ and $d_{\Re w,l}$. With these thresholds, the conditional probability of the real component can be evaluated as $P_{\mathsf{W}|\mathsf{U},\theta_e}\left(\Re{w}|\Re{ue^{j\theta_e[k]}}\right)=P(d_{\Re{\!w}\!,l}\le ue^{j\theta_e[k]} \le d_{\Re{\!w}\!,u})$, which can be computed using the $Q(\cdot)$ function. 
Inserting (\ref{eq:proboverpn}) into (\ref{eq:MUIforQAM}) the mutual information rate under phase noise can be evaluated numerically. Note that extensive computational resources are required for higher-order \gls{QAM} modulations, particularly as multiple phase noise variances $\sigma_\mathrm{PN}^2$ are required.

The  bandwidth is assumed not to be influenced by the phase noise and as the spectrum of the resulting  signal after filtering matches that of the
filter itself, the power-containment bandwidth of QAM is identical
to that of the used \gls{RRC} pulse.

\subsection{Information Rate of Time-Domain Modulation Schemes}

In contrast to \gls{QAM}, for which the mutual information can be evaluated on a symbol-by-symbol basis due to its memoryless structure, the time-domain characteristics of \gls{ZXM} and \gls{IR} introduce memory. Consequently, the mutual information must be computed over sequences. We employ the simulation-based auxiliary channel framework of \cite{arnold_simulation-based_2006} to estimate a lower bound. Its application to \gls{IR} and \gls{ZXM} is detailed in \cite{Roth202409} and \cite{Zeit202409}. The auxiliary channel lower bound is given by
\begin{equation}
I^\prime(\mathsf{U};\mathsf{W}) \ge H^\prime(\mathsf{U}) - \frac{1}{n}P^A_{\mathsf{W}}(\mathbf{w}) + \frac{1}{n}\log P^A_{\mathsf{U},\mathsf{W}}(\mathbf{u},\mathbf{w}),
\label{eq:muiZXM}
\end{equation}
where $P^A(\cdot)$ denotes the probability mass function of the auxiliary channel. This surrogate channel approximates the true system by imposing simplified memory assumptions and uncorrelated noise. The bound is evaluated by generating sufficiently long symbol sequences $\mathbf{u} = [u_1,\ldots,u_n]$ and $\mathbf{w} = [w_1,\ldots,w_n]$ corresponding to the transmitted and quantized received symbols, respectively. The required likelihoods of the symbol sequences are computed via the forward recursion of the \gls{BCJR} algorithm \cite{bahl1974optimal}.
For ZXM we adapt the auxiliary channel described in \cite{Zeit202409} by incorporating phase noise into the covariance structure in \cite[eq. (9)]{Zeit202409}, i.e., the covariance matrix $\boldsymbol{\Sigma}$ is extended to
\begin{equation}
\boldsymbol{\Sigma}=\boldsymbol{\Sigma}_\mathrm{AWGN}+\boldsymbol{\Sigma}_\mathrm{ISI}+\boldsymbol{\Sigma}_\mathrm{PN},\label{eq:PNStructureArnold}
\end{equation}
where  $\boldsymbol{\Sigma}_\mathrm{AWGN}$ and $\boldsymbol{\Sigma}_\mathrm{ISI}$ describe the effects of the \gls{AWGN} and inter-symbol-interference, respectively.
The entries of $\boldsymbol{\Sigma}_\mathrm{PN}$ are defined as
\begin{alignat}{2}
[\boldsymbol{\Sigma}_\mathrm{PN}]_{m,n} = \Phi_{ww}[m-n]\begin{cases}
2 \!-\! 2 e^{-\sigma_\mathrm{PN}^2/2},  &m=n\\
1\!+\!e^{-\sigma_\mathrm{PN}^2}\!-\!2 e^{-\sigma_\mathrm{PN}^2/2}, & m\neq n
\end{cases}  \nonumber
\end{alignat}
with autocorrelation $\Phi_{ww}[m-n]=\mathbb{E}[w_mw^*_n]$ of the phase-noise-free receive symbols $w$. 

Due to the considered energy detection for unipolar \gls{IR}, phase noise has no effect on the system performance, while for real-valued variable-sign configurations, phase noise leads to a reduced amplitude, i.e., the receiver only has access to $\Re{w e^{j\theta_e[k]}}$. Following the covariance structure in (\ref{eq:PNStructureArnold}), we model $\boldsymbol{\Sigma}_\mathrm{PN}$ for variable-sign \gls{IR} as
 \begin{alignat}{2}
 [\boldsymbol{\Sigma}_\mathrm{PN}]_{m,n}&\!=\!\Phi_{ww}[m\!-\!n]\begin{cases}
 \!\!\frac{1}{2} e^{-2\sigma_\mathrm{PN}^2}-2e^{-\sigma_\mathrm{PN}^2/2}+\frac{3}{2}, & m=n\\
 e^{-\sigma_\mathrm{PN}^2}-2e^{-\sigma_\mathrm{PN}^2/2}+1, & m\neq n.
 \end{cases} \nonumber
 \end{alignat}
The $99\%$ containment bandwidth for ZXM and IR is analyzed numerically. 
 \subsection{Phase-Noise Sensitivity Across Modulation Schemes}
\begin{figure}
\center
\resizebox{0.9\columnwidth}{!}{\input{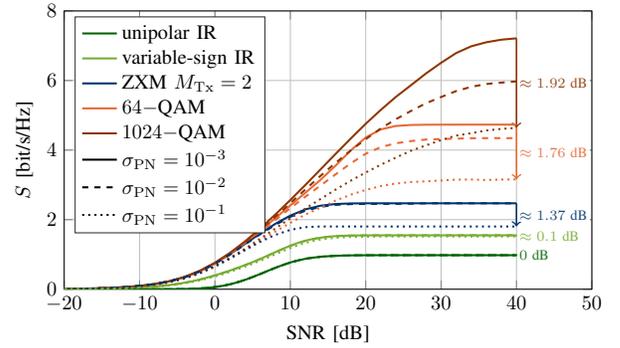}}
\caption{Spectral efficiency of exemplary modulation schemes}
\label{fig:SE_PN}
\end{figure}

The achievable spectral efficiency considering a $99\%$ power containment bandwidth for  an exemplary selection of modulation schemes from all three considered families is displayed in Fig.~\ref{fig:SE_PN} at exemplary levels of phase noise. This figure also contains the relative loss of spectral efficiency due to phase noise at $40\,\mathrm{dB}$ \gls{SNR}. 
It can be seen that modulation schemes with a high achievable spectral efficiency are more affected by phase noise and that \gls{IR} schemes are robust.
\section{Numerical Results}
\label{sec:numresults} 
Having modeled the dominant analog power contributions, the phase-noise trade-off, and its impact on spectral efficiency for different modulation schemes, we now evaluate the resulting system energy efficiency.
To this end, we perform a multi-stage optimization to quantify the energy-saving potential of the Gearbox-PHY. First, a design-time relaxation allows the analog front end to be optimized individually for each target rate, yielding an upper performance bound. Next, the front end parameters are fixed per gear to reflect implementable hardware. Finally, a cell-level optimization over a spatial user distribution is carried out to assess system-wide energy efficiency.\looseness-1

\subsection{Design-Time Optimization}
\label{subsec:variableOpt}
We begin with an optimization in which most front end parameters, especially, oscillator phase noise, are treated as continuous and variable for each target data rate $R_\mathrm{eff}$. This approach allows us to assess the theoretical performance limits before imposing practical restrictions.

Accordingly, we expand the initial optimization problem in (\ref{eq:OptimizationProblem}) with two more parameters, i.e.,
\begin{equation}
	   1\leq F\leq F_\mathrm{max},\qquad
	   \sigma_J\leq \sigma_{J,\mathrm{max}} \label{eq:OptimizationProblemVariableLO},
\end{equation}
to account for the (Wiener) phase noise process with variance $\sigma_J^2$ and pilot spacing $F$. To ensure periodic pilot tracking and a well-defined optimization problem, we cap the pilot spacing at $F_\mathrm{max}=10^5$, which also improves the numerical stability of the optimization. Moreover, to reflect the robustness of unipolar \gls{IR} while avoiding unrealistically low power consumption in (\ref{eq:VCO_Power_Jitter}), the phase noise standard deviation is limited to $\sigma_{J,\mathrm{max}}=\SI{0.5}{rad}$, preventing excessive phase noise that would otherwise invalidate the assumed constant signal bandwidth.

For all numerical results, we adopt the same parameter set, i.e., path loss exponent $\beta=2.5$, transmission distance $d=\SI{50}{m}$, antenna gain $D_\mathrm{R}=D_\mathrm{T}=\SI{6}{dB}$  (equal for receiver and transmitter and regardless of the carrier frequency), maximum transmit power $P_\mathrm{max}=\SI{10}{W}$, maximum bandwidth $B_\mathrm{max}=0.1 f_c$, transmitter sleep factor $\epsilon_\mathrm{Tx}=0.01$, receiver idle factor $\epsilon_\mathrm{Rx}=0.5$, \gls{LNA} gain $G_\mathrm{LNA}=\SI{15}{dB}$ \cite{LNASurveyPaper}, \gls{LNA} noise figure $N_\mathrm{LNA}=\SI{5}{dB}$ \cite{LNASurveyPaper}, bandwidth-dependent \gls{LNA} figure-of-merit $FOM_\mathrm{LNA}=10^{-7}$ \cite{Mezghani_PowerEfficiency}, \gls{DAC} steering current $I_0=\SI{10}{\mu A}$ \cite{CuiOptimization}, parasitic capacitance of \gls{DAC} $C_\mathrm{p}=\SI{1}{pF}$ \cite{CuiOptimization}, \gls{DAC} supply voltage $V_\mathrm{dd}=\SI{1}{V}$, local oscillator phase noise \gls{PSD} noise floor $K_\mathrm{0}=\SI{-125}{dBc/Hz}$, local oscillator operating temperature $\Gamma=\SI{300}{K}$, excess noise coefficient associated with MOS devices $\delta=1$ \cite{JitterPower}, reference oscillator power consumption $P_\mathrm{REF} = \SI{198}{\mu W}$ \cite{Crystal}, reference oscillator frequency $f_\mathrm{REF} = \SI{54}{MHz}$ \cite{Crystal}, charge pump and reference oscillator \gls{PSD} of $S_\mathrm{CP}=S_\mathrm{REF}=\SI{-160}{dBc/Hz}$ \cite{JitterPower,Crystal}.

Since the resulting optimization problem does not admit a closed-form solution, all results are obtained via numerical minimization. The optimization is performed separately for each target rate $R_\mathrm{eff}$ and jointly over the transmission bandwidth $B$, the duty cycle $\gamma$, the Wiener phase-noise variance $\sigma_J^2$, and the pilot spacing $F$. 
The required transmit power is determined implicitly from the SNR needed to achieve the corresponding spectral efficiency for the respective modulation scheme, which is defined by $R_\mathrm{eff}$, $\gamma$, and $B$.

\begin{figure}
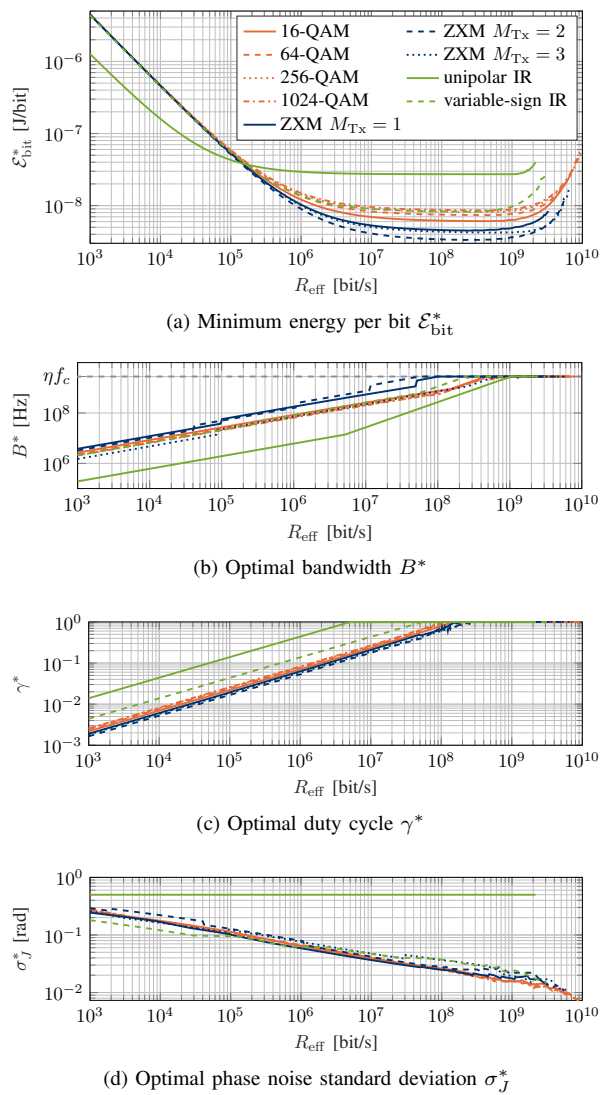
 
    \centering
  \subfloat[Minimum energy per bit $\mathcal{E}_\mathrm{bit}^*$\label{fig:VariableLO:EPerBit}]{%
       \resizebox{0.9\columnwidth}{!}{\input{figures/variableLO/Ebit}} }  
       \\
   \subfloat[Optimal bandwidth $B^*$\label{fig:VariableLO:BW}]{%
       \resizebox{0.9\columnwidth}{!}{\input{figures/variableLO/Bandwidth}} } 
       \\ 
    \subfloat[Optimal duty cycle $\gamma^*$\label{fig:VariableLO:gamma}]{%
       \resizebox{0.9\columnwidth}{!}{\input{figures/variableLO/Gamma}} }  
       \\ 
    \subfloat[Optimal phase noise standard deviation $\sigma_J^*$ \label{fig:VariableLO:sigma}]{%
       \resizebox{0.9\columnwidth}{!}{\input{figures/variableLO/Sigma}} } 
  \caption{Results of design-time numerical energy efficiency  optimization.}
  \label{fig:OptimizationVariableLO} 
\end{figure}
Fig.~\ref{fig:OptimizationVariableLO} shows the result of the numerical optimization exemplarily for $f_c=\SI{28}{GHz}$. As seen in Fig.~\ref{fig:VariableLO:EPerBit}, adapting the oscillator to the target data rate results in nearly identical minimum energy per bit $\mathcal{E}^*_\mathrm{bit}$ across all gears at low effective rates $R_\mathrm{eff}$, except for unipolar \gls{IR}, which operates without a downconversion stage at the receiver.

At low data rates, power consumption is dominated by the mixing stage, i.e., the \gls{LO} and mixer (not shown here). Although the oscillator power consumption can be reduced through adaptation, the mixer power is constant, according to Table~\ref{tab:MixerLOPow}, and remains the limiting factor. Consequently, the mixer power dictates the energy consumption across modulation schemes in this regime, leading to similar energy efficiencies. As $R_\mathrm{eff}^*$ increases, $\mathcal{E}_\mathrm{bit}^*$ first decreases, then saturates, and finally grows rapidly at high rates, driven by the transmit power required to reach maximum spectral efficiency.\looseness-1

The optimized bandwidth $B^*$ in Fig.~\ref{fig:VariableLO:BW} and duty cycle $\gamma^*$ in Fig.~\ref{fig:VariableLO:gamma} exhibit a similar trend. Both increase roughly linearly with the effective data rate $R_\mathrm{eff}$ in the logarithmic scale until their respective limits, $\gamma\le 1$ and $B \le \eta f_c$, are reached. Notably, unipolar \gls{IR} consistently operates with substantially lower bandwidth than the other modulation schemes.

Fig.~\ref{fig:VariableLO:sigma} 
highlights the behavior of the optimized Wiener phase noise standard deviation $\sigma_J^*$. As expected, unipolar \gls{IR}, which is assumed to be insensitive to phase noise, always selects the maximum admissible $\sigma_J$ imposed by (\ref{eq:OptimizationProblemVariableLO}). 
In contrast, all other modulation schemes favor progressively lower $\sigma_J^*$ as the effective rate $R_\mathrm{eff}$ increases, reflecting the growing importance of phase noise suppression for maintaining spectral efficiency at high rates. The resulting $\sigma_J^*$ values are largely comparable across modulation schemes, differing only marginally for a given $R_\mathrm{eff}$.
Interestingly, the pilot spacing is always driven to its upper bound $F^*=F_\mathrm{max}=10^5$ and is therefore omitted from the figures. This indicates that, with an adaptive oscillator, mitigating phase noise by directly improving oscillator quality is preferable to relying on pilot-assisted tracking in this design-time scenario.
Overall, for this idealized level of adaptivity, the differences in energy efficiency of the considered gears reach around one order of magnitude, see Fig.~\ref{fig:OptimizationVariableLO}. However, as such an \gls{LO} tuning with the required data rate is not feasible in practice, the present results should be interpreted as a theoretical baseline that motivates a more constrained optimization.

\subsection{Implementation-Constrained Optimization}
\label{subsec:ConstrOpt}
\begin{figure}
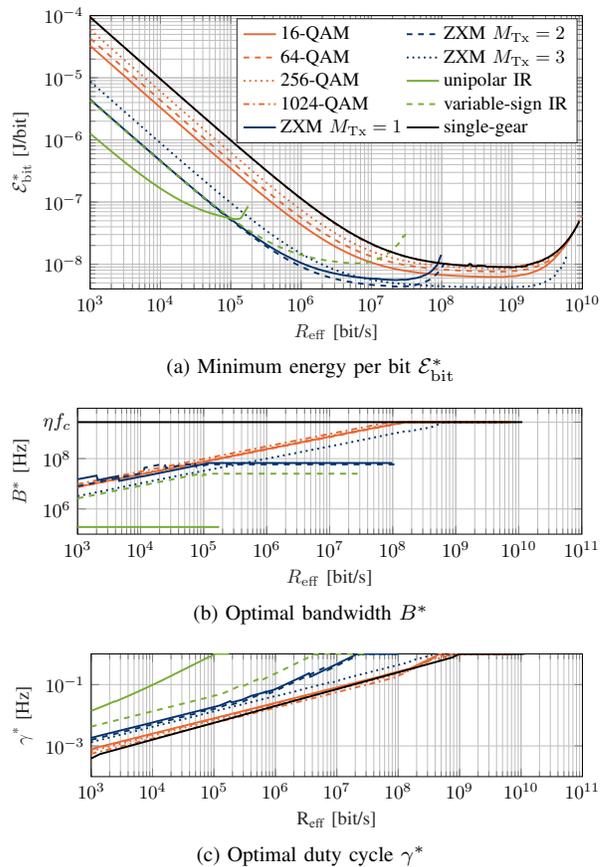

    \centering
  \subfloat[Minimum energy per bit $\mathcal{E}_\mathrm{bit}^*$\label{fig:fixedLO:EPerBit}]{%
       \resizebox{0.9\columnwidth}{!}{\input{figures/fixedLO/Ebit}} }  
       \\
   \subfloat[Optimal bandwidth $B^*$\label{fig:fixedLO:BW}]{%
       \resizebox{0.9\columnwidth}{!}{\input{figures/fixedLO/Bandwidth}} } 
       \\ 
    \subfloat[Optimal duty cycle $\gamma^*$\label{fig:fixedLO:gamma}]{%
       \resizebox{0.9\columnwidth}{!}{\input{figures/fixedLO/Gamma}} }  
   \caption{Results of energy efficiency optimization for fixed front ends.}
  \label{fig:OptimizationFixedLO} 
\end{figure}

Having established the impact of rate-dependent \gls{LO} selection, we now impose a practical implementation constraint by assuming that each gear employs a single, fixed analog front end. Although certain circuit blocks may allow runtime configurability, such as adjustments of the \gls{PA} operating point, the local oscillator and the low-noise amplifier are treated as non-adaptive components. In particular, phase-noise-tunable \glspl{VCO} are hard to realize in practice, and each gear is therefore assigned a dedicated \gls{LO} with a fixed phase noise standard deviation $\sigma_J$. Moreover, the bandwidth-dependent \gls{LNA} model used in the analysis represents a design-time abstraction, as the \gls{LNA} bandwidth is fixed after fabrication and consequently determines the maximum achievable system bandwidth $B_\mathrm{max}$. Since fixing these parameters changes the achievable optimum, we determine $\sigma_J^*$ and $B_\mathrm{max}^*\le \eta f_c$ via a staged procedure:
\begin{enumerate}
\item \textbf{Unconstrained front end:}
Solve (\ref{eq:OptimizationProblem}) over system parameters incl. (\ref{eq:OptimizationProblemVariableLO}) with variable \gls{LO} phase noise and \gls{LNA} bandwidth (cf. Section~\ref{subsec:variableOpt} and see Fig.~\ref{fig:OptimizationVariableLO}).
\item \textbf{Fix per-gear front end:}
Using $\mathcal{E}^*_\mathrm{bit}$ from Fig.~\ref{fig:VariableLO:EPerBit}, identify the rate interval where each gear is energy-optimal and fix $\sigma_J^*$ and $B_\mathrm{max}^*$ to their values at the lowest rate in that interval, as this is the most likely operating point, see \cite{Gast202403}. If a gear is never energy-optimal, its parameters are fixed at the rate $R_\mathrm{eff}$ where the gap to the minimum achievable energy per bit (attained by another gear) is smallest. The resulting $\sigma_J^*$  for $f_c=\SI{28}{GHz}$ are summarized in Tab.~\ref{tab:jitter}, while the corresponding $B^*_\mathrm{max}$ values are shown later in Fig.~\ref{fig:fixedLO:BW}.

\begin{table}
\centering
\caption{Resulting Wiener phase noise variances after variable optimization for $f_c=\SI{28}{GHz}$\vspace{-6pt}}
\renewcommand{\arraystretch}{1.2}
\setlength{\tabcolsep}{1.8pt}
\begin{tabular}{|c|c|c|c|c|c|c|c|c|c|} \hline
\multirow{2}{*}{gear} 
& \multicolumn{4}{c|}{QAM ($M$)} 
& \multicolumn{3}{c|}{ZXM ($M_\mathrm{Tx}$)} 
& \multicolumn{2}{c|}{IR} \\ \cline{2-10}
& $16$ & $64$ & $256$ & $1024$ & $1$ & $2$ & $3$ & Unipolar & Variable \\ \hline
$\sigma_J^*$ 
& $0.01$ 
& $0.0093$
& $0.0085$
& $0.0075$
& $0.093$
& $0.12$
& $0.0017$
& $0.5$
& $0.0093$ \\ \hline
\end{tabular}
\label{tab:jitter}
\end{table}

\item \textbf{Re-optimize:}
Resolve the initial optimization problem (\ref{eq:OptimizationProblem}) with $\sigma_J^*$ and $B_\mathrm{max}^*$ fixed for each gear.\looseness-1
\end{enumerate}

Fig.~\ref{fig:OptimizationFixedLO} visualizes the solution of the proposed multi-stage optimization, again for $f_c=\SI{28}{GHz}$, and also includes a single-gear reference, assumed to use the same fixed oscillator as used for $1024$-QAM, while operating strictly at the maximum available bandwidth $B=\eta f_c$. This is chosen as a representation for conventional systems, using a fixed channel bandwidth.\looseness-1

Several observations follow. Firstly, the reduced maximum bandwidth in Fig.~\ref{fig:fixedLO:BW} causes low-order gears to terminate at lower data rates than in the variable front end case in Fig.~\ref{fig:OptimizationVariableLO}, further highlighting that each gear is only optimal over a limited rate interval. As already indicated in Fig.~\ref{fig:VariableLO:EPerBit}, \gls{ZXM} with $M_\mathrm{Tx}=2$ remains energy-optimal over a broad range of rates, which leads to the selection of a comparatively small $B_\mathrm{max}$ and a large oscillator phase noise standard deviation, see Tab.~\ref{tab:jitter}.
Surprisingly, like in the variable front end case, the optimal pilot spacing satisfies $F^*=F_\mathrm{max}$ for all rates. This is explained by the fact that phase-noise-sensitive schemes, in particular high-order \gls{QAM}, are only energy-efficient at high data rates, where they rely on the chosen high-quality oscillators and do not sacrifice rate for additional phase noise tracking.\looseness-1

Finally, the single-gear solution is nearly indistinguishable from adaptive $1024$-QAM in terms of $\mathcal{E}^*_\mathrm{bit}$, confirming that bandwidth adaptation alone provides limited energy gains. In contrast, modulation schemes with fundamentally different front ends yield substantial improvements. E.g., at low $R_\mathrm{eff}$, \gls{IR} schemes achieve energy savings of up to three orders of magnitude for the case of $f_c=\SI{60}{GHz}$ (not shown here but shown in \cite{Gast202503}) compared to the single-gear approach. These observed energy reductions are significant and underscore the potential of the Gearbox-PHY for energy-efficient operation. At the same time, the outcome of the multi-stage optimization strongly depends on the parameters, and especially on the communication distance $d$.
Here, the evaluation is restricted to a single link distance of $d=\SI{50}{m}$, which does not reflect the range of operating conditions in real-world scenarios.
\subsection{Cell-based Evaluation}
As a last step, we extend the analysis to a cellular perspective by assuming  users to be spread over a cell, leading to varying communication distances and  path losses. We therefore evaluate the energy per bit over distances $d_\mathrm{min}\leq d\leq d_\mathrm{max}$, while retaining a single fixed \gls{LO} and \gls{LNA} per gear.

Since front end parameters cannot be optimized for all distances simultaneously, we assume users to be uniformly distributed over the cell area, implying a quadratic increase in user count with distance. Based on this assumption, a median distance $d_\mathrm{m}$ is defined such that half of the users are closer than $d_\mathrm{m}$ and half are farther away, 
yielding the design point for fixing the front end parameters, which is done following the method described in Section~\ref{subsec:ConstrOpt}. 

With the front end optimized for the median distance $d_\mathrm{m}$, we apply the multi-stage optimization while evaluating performance over the full cell. For this, the Gearbox-PHY is analyzed at discrete distances $d_i\in[d_\mathrm{min},d_\mathrm{max}]$ with step size $\Delta_d$, capturing distance-dependent path loss variations.

Assuming a uniform user distribution over the cell area, we compute an area-weighted energy-per-bit metric as
\begin{equation} \mathcal{E}^*_\mathrm{bit,Area}=\frac{1}{A_\mathrm{cell}}\sum_i \mathcal{E}^*_\mathrm{bit}(d_i) \cdot 2 \pi d_i \Delta_d, \end{equation}
where $A_\mathrm{cell}$ is the total cell area and $2\pi d_i \Delta_d$ corresponds to the annular area at distance $d_i$ and, thus, the user fraction at $d_i$.\looseness-1

\begin{figure}
    \centering
    \resizebox{0.9\columnwidth}{!}{\input{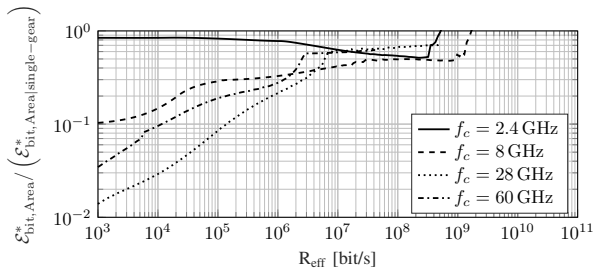}}
    \caption{Weighted savings of Gearbox-PHY over cell with minimum distance $d_\mathrm{min}=\SI{20}{m}$, maximum distance $d_\mathrm{max}=\SI{400}{m}$, ring width $\Delta_d=\SI{40}{m}$}
    \label{fig:Impairments:Area400}
\end{figure}

The achievable energy savings of such an optimization, given as the energy per bit of the Gearbox approach $\mathcal{E}^*_\mathrm{bit,Area}$, by selecting the optimal gear for each rate, relative to the energy per bit achievable with a single-gear $\mathcal{E}^*_\mathrm{bit,Area|single-gear}$ is shown in Fig.~\ref{fig:Impairments:Area400} for a cell radius of $\SI{400}{m}$.
While the highest gains occur at short ranges (e.g., nearly two orders of magnitude for $f_c=\SI{28}{GHz}$ at $d=\SI{50}{m}$, see Fig.~\ref{fig:fixedLO:EPerBit}), increased path loss at larger distances reduces these benefits. Nevertheless, significant energy savings  of up to two orders of magnitude persist, particularly at low data rates and higher carrier frequencies, i.e., $f_c=\SI{28}{GHz}$ and $f_c=\SI{60}{GHz}$. In contrast, at $f_c=\SI{2.4}{GHz}$, the achievable savings are marginal, indicating that high spectrally efficient modulation is near energy-optimal at low carrier frequencies.\looseness-1

\section{Conclusion}
\label{sec:Conclusion}
In this work, the Gearbox-PHY was investigated as an energy-centric alternative to conventional single-gear transceivers that are predominantly optimized for peak-rate operation. Guided by the high likelihood of low data rates, we showed that bandwidth scaling and duty cycling alone are insufficient to maximize energy efficiency gains. Instead, adaptivity in the analog front is needed.

To move beyond idealized assumptions, we incorporated analog hardware impairments, specifically phase noise and finite quantizer resolution, and jointly modeled their impact on front end power consumption and spectral efficiency. Accounting for these non-idealities leads to a significant saving potential, as the robustness of low-complexity modulation schemes w.r.t. distortions can be exploited to increase energy efficiency.

To assess implementable operation, we fixed one analog front end per gear. Even under these constraints, substantial energy reductions remain achievable. In a cellular evaluation with users distributed over a cell area, the Gearbox-PHY maintains significant gains at low data rates, which are the most probable operating points, see \cite{Gast202403}, and high carrier frequencies. Conversely, at low carrier frequencies the advantage becomes marginal, consistent with the observation that conventional highly spectrally-efficient \gls{QAM} remains energy-optimal in that regime.

These results are directly relevant for future networks, as 6G must improve energy efficiency per delivered bit by approximately one to two orders of magnitude to offset the projected traffic growth \cite{EricssonMobilityReport2024, Gearbox}. Our results show that such gains are achievable with the Gearbox-PHY even in a cellular setting that accounts for hardware impairments, making it a key building block for energy-aware PHY design. 
Beyond traffic growth, improving energy efficiency remains the primary lever for reducing operational expenditures.

Future work should focus on extending the Gearbox-PHY framework to multi-user, multi-antenna scenarios, additional hardware impairments, most notably power-amplifier nonlinearities, and on integrating digital front end power consumption into the optimization. Moreover, validating the concept through prototype implementations will be essential to assess its practical feasibility and impact in real 6G deployments.
\bibliographystyle{IEEEtran}
\bibliography{references}

@INPROCEEDINGS{Gast202205,
  author={{F.} {Gast} and M. Schlüter and M. Dörpinghaus and H. Halbauer and G. Fettweis},
  title={Phase Noise Tracking for Receivers with 1-bit Quantization and Oversampling},
  booktitle={Proc. IEEE Int. Conf. Commun. (ICC)},
  month={May.},
  year={2022},
  address={Seoul, Korea (South)},
}

@INPROCEEDINGS{Zeit202212,
  author={S. Zeitz and {F.} {Gast} and M. Dörpinghaus and G. Fettweis},
  title={On the {Bayesian} {Cramér-Rao} Bound for Phase Noise Estimation Based on 1-bit Quantized Samples},
  booktitle={Proc. IEEE Glob. Commun. Conf. (GLOBECOM)},
  month={Dec.},
  year={2022},
  address={Rio de Janeiro, Brazil},
}

@INPROCEEDINGS{Gast202307,
  author={{F.} {Gast} and S. Zeitz and M. Dörpinghaus and G. Fettweis},
  title={Comparing Iterative and Least-Squares Based Phase Noise Tracking in Receivers with 1-bit Quantization and Oversampling},
  booktitle={Proc. IEEE Stat. Signal Process. Workshop (SSP)},
  month={Jul.},
  year={2023},
  address={Hanoi, Vietnam},
}

@INPROCEEDINGS{Gast202403,
  author={{F.} {Gast} and M. Dörpinghaus and F. Roth and G. Fettweis},
  title={A New Spatio-Temporal Model for Data Rate Distributions in Mobile Networks},
  booktitle={Proc. Int. Workshop Smart Antennas (WSA)},
  month={Mar.},
  year={2024},
  address={Dresden, Germany},
}

@ARTICLE{Gast202404,
  author={{F.} {Gast} and M. Dörpinghaus and P. Sen and A. Nimr and G. Fettweis},
  title={Hardware-Aware Energy Efficiency Optimization in Wireless Communications using a {Gearbox-PHY}},
  journal={IEEE Commun. Lett.},
  month={Apr.},
  year={2024},
  doi={10.1109/LCOMM.2024.3395759},
}

@INPROCEEDINGS{Gast202407,
  author={{F.} {Gast} and F. Roth and M. Dörpinghaus and P. Sen and S. Zeitz and G. Fettweis},
  title={Energy Optimization using Joint Modulation Scheme and Front End Adaptation - the {Gearbox-PHY}},
  booktitle={Proc. Int. Symp. Wireless Commun. Syst. (ISWCS)},
  month={Jul.},
  year={2024},
  address={Rio de Janeiro, Brazil},
}

@INPROCEEDINGS{Roth202409,
  author={F. Roth and M. Dörpinghaus and S. Zeitz and {F.} {Gast} and G. Fettweis},
  title={Why to Use the Phase in Time-Encoding Modulation and Its Effect on the Spectral Efficiency},
  booktitle={Proc. IEEE Int. Symp. Pers., Indoor Mobile Radio Commun. (PIMRC)},
  month={Sep.},
  year={2024},
  address={Valencia, Spain},
}

@INPROCEEDINGS{Zeit202409,
  author={S. Zeitz and F. Roth and {F.} {Gast} and M. Dörpinghaus and G. Fettweis},
  title={Timing Synchronization and Detection for Systems with 1-bit Quantization and Runlength Coding},
  booktitle={Proc. IEEE Int. Workshop Signal Process. Adv. Wireless Commun. (SPAWC)},
  month={Sep.},
  year={2024},
  address={Lucca, Italy},
  doi={https://doi.org/10.1109/SPAWC60668.2024.10694337},
}

@ARTICLE{Meller202411,
  author={G. Meller and {F.} {Gast} and F. Protze and J. Wagner and F. Ellinger and G. Fettweis},
  journal={IEEE Trans. Microw. Theory Techn.}, 
  title={Noise Analysis of a 434-{MHz} Wakeup Receiver Analog Frontend Core With  $-$ 93-{dBm} Input Sensitivity and 65-{pJ}/Bit Efficiency Based on a Switched Injection-Triggered Oscillator With Surface Acoustic Wave Resonator}, 
  year={2024},
  pages={1-16},
  doi={10.1109/TMTT.2024.3482456}}

@INPROCEEDINGS{Gast202503,
  author={{F.} {Gast} and F. Roth and M. Dörpinghaus and P. Sen and S. Zeitz and G. Fettweis},
  title={The Role of Oscillator Phase Noise in Maximizing Transceiver Energy Efficiency},
  booktitle={Proc. IEEE Wireless Commun. Netw. Conf. (WCNC)},
  month={Mar.},
  year={2025},
  address={Milan, Italy}
}

@INPROCEEDINGS{Gast202503b,
  author={{F.} {Gast} and K. Xu and M. Dörpinghaus and G. Fettweis},
  title={Improving Power Amplifier Efficiency of Zero-Crossing Modulation at Sub-{THz} Frequencies},
  booktitle={Proc. Ger. Microw. Conf. (GeMiC)},
  month={Mar.},
  year={2025},
  address={Dresden, Germany}
}

@INCOLLECTION{6GlifeBook,
title = {Chapter 6 - {6G} energy-efficient physical layer},
editor = {Frank H.P. Fitzek and Holger Boche and Wolfgang Kellerer and Patrick Seeling},
booktitle = {6G-life},
publisher = {Academic Press},
pages = {89-118},
year = {2026},
doi = {https://doi.org/10.1016/B978-0-44-327410-7.00018-1},
author = {Florian {Gast et al.}},
}

@article{NGMN2021green,
  title={Green future networks: Network energy efficiency},
  author={{{NGMN Alliance}}},
  journal={NGMN, Whitepaper},
  year={2021}
}

@article{NGMN2021drivers,
  title={{6G} Drivers and Vision},
  author={{{NGMN Alliance}}},
  journal={NGMN, Whitepaper},
  year={2021}
}

@inproceedings{sharma2013issues,
  title={Issues and challenges in wireless sensor networks},
  author={Sharma, S. and Bansal, R. K. and Bansal, S.},
  booktitle={Proc. Int. Conf. Mach. Intell. Res. Adv.},
  pages={58--62},
  year={2013},
  address={Katra, India},
  month={Oct.}
}

@article{malmodin2018energy,
  title={The Energy and Carbon Footprint of the Global {{ICT}} and {{E\&M}} Sectors 2010--2015},
  author={Malmodin, J. and Lund{\'e}n, D.},
  journal={Sustainability},
  volume={10},
  number={9},
  pages={3027},
  year={2018},
  publisher={Multidisciplinary Digital Publishing Institute}
}

@misc{GSMA2019,
    author={GSMA},
    title={2019 Mobile Industry Impact Report: Sustainable Development Goals}

}

@misc{GSMA2021,
    author={GSMA},
    title={The Enablement Effect},
    year = {2021}

}

@misc{GSMA2021_SustainableTelco,
    author={{{GSMA}}intelligence},
    title={Radar: The sustainable telco},
    year = {2021}

}

@misc{GSMA2021_Benchmarking,
    author={{{GSMA}}intelligence},
    title={Going green: benchmarking the energy efficiency of mobile},
    year = {2021}

}

@techreport{bles_umweltbezogene_nodate,
  title = {Umweltbezogene {Technikfolgenabschätzung} {Mobilfunk} in {Deutschland}},
  author = {Stobbe et al., L. },
  langid = {german},
  year={2023},
  institution= {Fraunhofer-Institut für Zuverlässigkeit und Mikrointegration, IZM}
}

@misc{5GGuide,
    author={GSMA},
    title={The 5{G} Guide - A Reference for Operators},
    year = {2019}

}

@misc{EricssonMobilityReport2024,
    author={Ericsson},
    title={Mobility Report},
    year = {2024}
}

@ARTICLE{Gearbox,

  author={Fettweis, G. P. and Boche, H.},

  journal={IEEE BITS Inf. Theory Mag.}, 

  title={6{G}: The Personal Tactile Internet—And Open Questions for Information Theory}, 

  year={2021},

  volume={1},

  number={1},

  pages={71-82},
  doi={10.1109/MBITS.2021.3118662}}

@ARTICLE{SpatialTraffic14,

  author={Lee, D. and Zhou, S. and Zhong, X. and Niu, Z. and Zhou, X. and Zhang, H.},

  journal={IEEE Trans. Wireless Commun.}, 
month={Feb.},
  title={Spatial modeling of the traffic density in cellular networks}, 

  year={2014},

  volume={21},

  number={1},

  pages={80-88},

  doi={10.1109/MWC.2014.6757900}}

@ARTICLE{OkumuraHata,

  author={Hata, M.},

  journal={IEEE Trans. Veh. Technol.}, 

  title={Empirical formula for propagation loss in land mobile radio services}, 

  year={1980},

  volume={29},

  number={3},

  pages={317-325},

  doi={10.1109/T-VT.1980.23859}
}

@misc{adcsurvey,
   author = {Murmann, B.},
   title = {{ADC Performance Survey 1997-2023}},
   note = {[Online]. Available: \url{https://github.com/bmurmann/ADC-survey}}
}

@INPROCEEDINGS{ADCSurveyKeynotePaper,
  author={B. Murmann},
  booktitle={Proc. IEEE Faible Tension Faible Consomm. (FTFC)},
  title={Energy Limits in {A}/{D} Converters},
  year={2013},
  month={Jun.},
  address={Paris, France},
  pages={1--4},
  doi={10.1109/FTFC.2013.6577781}
}

@article{DACSurvey,
  title={A survey of high-speed high-resolution current steering {DACs}},
  author={X. Li and L. Zhou},
  journal={J. Semicond.},
  volume={41},
  number={11},
  pages={111404},
  year={2020},
  publisher={IOP Publishing}
}

@article{bossy2020flexible,
  title={Flexible, Brain-Inspired Communication in Massive Wireless Networks},
  author={B. Bossy and P. Kryszkiewicz and H. Bogucka},
  journal={Sensors},
  volume={20},
  number={6},
  pages={1587},
  year={2020},
  publisher={MDPI}
}

@ARTICLE{CuiOptimization,

   author={S. Cui and A. J. Goldsmith and A. Bahai},

  journal={IEEE Trans. Wireless Commun.}, 

  title={Energy-constrained modulation optimization}, 

  year={2005},

  volume={4},

  number={5},

  pages={2349-2360},

  doi={10.1109/TWC.2005.853882}}

@INPROCEEDINGS{MuradMixer2GHz,
  author={S. A. Z. Murad and S. N. Mohyar and A. Harun and M. N. M. Yasin and I. S. Ishak and R. Sapawi},
  booktitle={Proc. IEEE Student Conf. Res. Develop. (SCOReD)},
  title={Low Noise Figure 2.4 {G}{H}z Down Conversion {C}{M}{O}{S} Mixer for Wireless Sensor Network Application},
  year={2016},
  month={Dec.},
  address={Kuala Lumpur, Malaysia},
  doi={10.1109/SCORED.2016.7810032}
}

@article{chang202128,
  title={A 28-{G}{H}z bidirectional active {G}ilbert-cell mixer in 90-nm {C}{M}{O}{S}},
  author={Y.-T. Chang and K.-Y. Lin},
  journal={IEEE Microw. Wireless Compon. Lett.},
  volume={31},
  number={5},
  pages={473--476},
  year={2021},
  publisher={IEEE}
}

@article{lee201360,
  title={60 {G}{H}z {C}{M}{O}{S} downconversion mixer with 15.46 d{B} gain and 64.7 d{B} {L}{O}-{R}{F} isolation},
  author={Lee, J.H. and Lin, Y.S.},
  journal={Electron. Lett.},
  volume={49},
  number={4},
  year={2013},
  publisher={Wiley Online Library}
}

@inproceedings{mahmoudi20048,
  title={8 {GHz}, 1{V}, high linearity, low power {CMOS} active mixer},
  author={F. Mahmoudi and C. A. T. Salama},
  booktitle={Proc. IEE Radio Freq. Integr. Circuits Syst., Dig. Papers },
  pages={401--404},
  year={2004},
  month={Aug.},
  address={Fort Worth, TX, USA}
}

@article{zhang2022optimal,
  title={Optimal design of {C}{M}{O}{S} mixer: A research review},
  author={H. Zhang and S. Tang and M. Cai and Y. Jiang},
  journal={Int. J. {R}{F} Microw. Comput.-Aided Eng.},
  volume={32},
  number={12},
  year={2022},
  publisher={Wiley Online Library}
}

@ARTICLE{JitterPower,

  author={Razavi, B.},

  journal={IEEE Trans. Circuits and Syst. I: Regul. Pap.}, 

  title={Jitter-Power Trade-Offs in {{PLL}}s}, 

  year={2021},

  volume={68},

  number={4},

  pages={1381-1387},

  doi={10.1109/TCSI.2021.3057580}}

@misc{PASurvey,
  author = {H. Wang and T.-Y. Huang and N. S. Mannem and J. Lee and E. Garay and D. Munzer and E. Liu and Y. Liu and B. Lin and M. Eleraky and H. Jalili and J. Park and S. Li and F. Wang and A. S. Ahmed and C. Snyder and S. Lee and H. T. Nguyen and M. E. D. Smith},
  title = {Power Amplifiers Performance Survey 2000--Present},
  date = {},
  note = {}
}

@INPROCEEDINGS{Mezghani_PowerEfficiency,

  author={A. Mezghani and J. A. Nossek},
  booktitle={Proc. Int. ITG Workshop Smart Antennas (WSA)},

  title={Modeling and minimization of transceiver power consumption in wireless networks}, 

  year={2011},

address={Aachen, Germany },
month={Feb.},

  volume={},

  number={},

  doi={10.1109/WSA.2011.5741951}}

@ARTICLE{LNASurveyPaper,

  author={L. Belostotski and S. Jagtap},

  journal={IEEE Solid-State Circuits Mag.}, 

  title={Down With Noise: An Introduction to a Low-Noise Amplifier Survey}, 

  year={2020},

  volume={12},

  number={2},

  pages={23-29},

  doi={10.1109/MSSC.2020.2987505}}

@ARTICLE{DitheredQuant,

  author={Gray, R. M. and Stockham, T. G.},

  journal={IEEE Trans. Inf. Theory}, 

  title={Dithered quantizers}, 

  year={1993},

  volume={39},

  number={3},

  pages={805-812},

  doi={10.1109/18.256489}}

@INPROCEEDINGS{ZXM,

  author={G. {Fettweis} and M. {Dörpinghaus} and S. {Bender} and L. {Landau} and P. {Neuhaus} and M. {Schlüter}},

  booktitle={Proc. 53rd Asilomar Conf. Signals, Syst. Comput.}, 

  title={Zero Crossing Modulation for Communication with Temporally Oversampled 1-Bit Quantization}, 

  year={2019},
  month={Nov.},
  address={ Pacific Grove, CA, USA},
  volume={},

  number={}}

@article{neuhaus2021zero,
  title={Zero-crossing modulation for wideband systems employing 1-bit quantization and temporal oversampling: Transceiver design and performance evaluation},
  author={Neuhaus, P. and D{\"o}rpinghaus, M. and Fettweis, G.},
  journal={IEEE Open J. Commun. Soc.},
  volume={2},
  pages={1915--1934},
  year={2021},
  publisher={IEEE}
}

@article{ziabakhsh2018peak,
  title={The peak-{SNR} performances of voltage-mode versus time-mode circuits},
  author={S. Ziabakhsh and G. Gagnon and G. W. Roberts},
  journal={IEEE Trans. Circuits Syst. II, Exp. Briefs},
  volume={65},
  number={12},
  pages={1869--1873},
  year={2018}
}

@book{wang2015rf,
  title={{RF} and mm-wave Power Generation in Silicon},
  author={H. Wang and K. Sengupta},
  year={2015},
  publisher={Academic Press}
}

@INPROCEEDINGS{PAPR,


    title={Root-raised cosine filter influences on {{PAPR}} distribution of single carrier signals},
  author={S. Daumont and B. Rihawi and Y. Lout},
  booktitle={Proc. 3rd Int. Symp. Commun., Control Signal Process. (ISCCSP)},
  pages={841--845},
address={ Saint Julian's, Malta },
month={Mar.},
  year={2008},

  doi={10.1109/ISCCSP.2008.4537340}}

@inproceedings{wolf2010maximum,
  title={On the Maximum Efficiency of Power Amplifiers in {{OFDM}} Broadcast Systems With Envelope Following},
  author={R. Wolf and F. Ellinger and R. Eickhoff},
  booktitle={Proc. Int. ICST Conf. Mobile Lightw. Wireless Syst. (MOBILIGHT)},
month={May.},
address={Barcelona, Spain},
  pages={160--170},
  year={2010},
  organization={Springer}
}

@INPROCEEDINGS{neuhaus2021EE,

  author={P. Neuhaus and M. Schlüter and C. Jans and M. Dörpinghaus and G. Fettweis},

  booktitle={Proc. 2021 Joint Eur. Conf. on Networks and Commun. \& 6G Summit (EuCNC/6G Summit)}, 

  title={Enabling Energy-Efficient {T}bit/s Communications by 1-Bit Quantization and Oversampling}, 

  year={2021},
  month={Jul.},
address={Porto, Portugal},

  doi={10.1109/EuCNC/6GSummit51104.2021.9482427}}

@INPROCEEDINGS{PowerDetector,

  author={K. Kang and P. D. Dong and J. Brinkhoff and C.-H. Heng and F. Lin and X. Yuan},
  booktitle={Proc. IEEE Int. Symp. Radio-Freq. Integr. Technol. (RFIT)},

  title={A power efficient 60 {{GHz}} 90nm {{CMOS}} {{OOK}} receiver with an on-chip antenna}, 

  year={2009},

  volume={},

  number={},
address={Singapore},
month={Jan.},
pages={},

  doi={10.1109/RFIT.2009.5383682}}

@ARTICLE{Goldsmith_ACM,

  author={Goldsmith, A.J. and Chua, S.-G.},

  journal={IEEE Trans. Commun.}, 

  title={Adaptive coded modulation for fading channels}, 

  year={1998},

  volume={46},

  number={5},

  pages={595-602},

  doi={10.1109/26.668727}}

@misc{ericsson2023ai,
  author       = {Elena Fersman and Johan Pettersson and Anette Höglund and Erik Sanders and Lackis Eleftheriadis},
  title        = {Intelligent sustainability: the role of {AI} in energy consumption, management and new revenues},
  howpublished = {Ericsson Blog},
  month        = jun,
  year         = {2023},
  note         = {Accessed: 2025-07-07},
}

@ARTICLE{arnold_simulation-based_2006,
  title = {Simulation-Based Computation of Information Rates for Channels With Memory},
  volume = {52},
  issn = {1557-9654},
  doi = {10.1109/TIT.2006.878110},
  pages = {3498--3508},
  number = {8},
  journal = {{IEEE} Trans. Inf. Theory},
  author = {Arnold, D. M. and Loeliger, H.-A. and Vontobel, P.O. and Kavcic, A. and Zeng, W.},
  year= {2006}}

@article{bahl1974optimal,
  title={Optimal decoding of linear codes for minimizing symbol error rate (corresp.)},
  author={L. Bahl and J. Cocke and F. Jelinek and J. Raviv},
  journal={IEEE Trans. Inf. Theory},
  volume={20},
  number={2},
  pages={284--287},
  year={1974},
  publisher={IEEE}
}

@ARTICLE{KhanzadiPhaseNoiseModel,

  author={M. R. Khanzadi and D. Kuylenstierna and A. Panahi and T. Eriksson and H. Zirath},
  journal={IEEE Trans. Circuits Syst. I, Regul. Pap.}, 

  title={Calculation of the Performance of Communication Systems From Measured Oscillator Phase Noise}, 

  year={2014},

  volume={61},

  number={5},
  pages={1553-1565},

  doi={10.1109/TCSI.2013.2285698}}

@INPROCEEDINGS{SimonPhaseNoise,

  author={Simon, V. and Senst, A. and Speth, M. and Meyr, H.},

  booktitle={Proc. IEEE Glob. Telecommun. Conf. (GLOBECOM)},

  title={Phase Noise Estimation via Adapted Interpolation},

  year={2001},
  address={San Antonio, TX, USA},
  month={Nov.},
  volume={6},

  number={},
pages={},
  pages={3297-3301 vol.6},

  doi={10.1109/GLOCOM.2001.966296}}

@ARTICLE{LS,

  author={M. Schlüter and M. Dörpinghaus and G. P. Fettweis},

  journal={IEEE Trans. Commun.}, 

  title={Joint Phase and Timing Estimation With 1-Bit Quantization and Oversampling}, 

  year={2022},

  volume={70},

  number={1},
  
  month={Jan.},

  pages={71-86},

  doi={10.1109/TCOMM.2021.3113946}}

@INPROCEEDINGS{Szortyka2024PLL,

  author={V. Szortyka and Q. Shi and K. Raczkowski and B. Parvais and M. Kuijk and P. Wambacq},

   booktitle={Proc. IEEE Int. Solid-State Circuits Conf. - Dig. Tech. Papers (ISSCC)},

  title={21.4 A 42{mW} 230fs-jitter sub-sampling 60{GHz} {PLL} in 40nm {CMOS}}, 

  year={2014},

  volume={},

  number={},

  pages={366-367},
  month={Feb.},
  address={San Francisco, CA, USA},

  doi={10.1109/ISSCC.2014.6757472}}

@book{meyr1998digital,
  title={Digital communication receivers: synchronization, channel estimation and signal processing},
  author={H. Meyr and M. Moeneclaey and S. A. Fechtel},
  year={1998},
  publisher={Wiley}
}

@ARTICLE{doerpinghaus2012Joint,

  author={D{\"o}rpinghaus, M. and Ispas, A. and Meyr, H.},

  journal={IEEE Trans. Inf. Theory}, 

  title={On the Gain of Joint Processing of Pilot and Data Symbols in Stationary {R}ayleigh Fading Channels}, 

  year={2012},

  volume={58},

  number={5},
pages={},
  pages={2963-2982},

  doi={10.1109/TIT.2011.2181818}}

@inproceedings{li2018200,
  title={A 200-{GHz} sub-harmonic injection-locked oscillator with 0-{dBm} output power and 3.5\% {DC}-to-{RF}-efficiency},
  author={Li, S. and Fritsche, D. and Carta, C. and Ellinger, F.},
  booktitle={Proc. IEEE Radio Freq. Integr. Circuits Symp. (RFIC)},
  pages={212--215},
  year={2018},
  month={Jun.},
  address={Philadelphia, PA, USA},
}

@inproceedings{laemmle2013fully,
  title={A Fully Integrated 120-{GHz} Six-Port Receiver Front-End in a 130-nm {SiGe} {BiCMOS} Technology},
  author={B. Laemmle and K. Schmalz and J. Borngraeber and J. C. Scheytt and R. Weigel and A. Koelpin and D. Kissinger},
  booktitle={Proc. IEEE Topical Meet. Silicon Monolithic Integr. Circuits RF Syst. (SiRF)},
  pages={129--131},
  year={2013},
  month={Jan.},
  address={Austin, TX, USA},
  doi={10.1109/SiRF.2013.6489455}
}

@ARTICLE{Crystal,

  author={K. M. Megawer and N. Pal and A. Elkholy and M. G. Ahmed and A. Khashaba and D. Griffith and P. K. Hanumolu},
  
  journal={IEEE J. Solid-State Circuits}, 

  title={A Fast Startup {{CMOS}} Crystal Oscillator Using Two-Step Injection}, 

  year={2019},

  volume={54},

  number={12},
pages={},
  pages={3257-3268},

  doi={10.1109/JSSC.2019.2936296}}

@ARTICLE{LNAFoMs,

   author={L. Belostotski and E. A. M. Klumperink},
  journal={IEEE Trans. Circuits Syst. II, Exp. Briefs},

  title={Figures of Merit for {CMOS} Low-Noise Amplifiers and Estimates for Their Theoretical Limits}, 

  year={2022},

  volume={69},

  number={3},

  pages={734-738},

  doi={10.1109/TCSII.2021.3113607}}

\end{document}